\documentclass[fleqn,usenatbib]{mnras}

\usepackage{newtxtext,newtxmath}

\usepackage[T1]{fontenc}

\DeclareRobustCommand{\VAN}[3]{#2}
\let\VANthebibliography\thebibliography
\def\thebibliography{\DeclareRobustCommand{\VAN}[3]{##3}\VANthebibliography}

\usepackage{graphicx}	
\usepackage{amsmath}	
\usepackage{fontawesome}
\usepackage{multirow}
\usepackage{xcolor}
\usepackage[normalem]{ulem}
\usepackage{cuted}

\newcommand{\bfk}{\mbox{\boldmath$k$}}

\newcommand{\be}{\begin{equation}}
\newcommand{\ee}{\end{equation}}
\newcommand{\bea}{\begin{eqnarray}}
\newcommand{\eea}{\end{eqnarray}}

\newcommand{\HiCOLA}{\texttt{Hi-COLA}~}

\newcommand{\betahc}{$\beta_{\textrm{HC}}$}
\newcommand{\shc}{$S_{\textrm{HC}}$}

\newcommand{\HH}{ \mathcal{H}}

\title[]{Matter Power Spectra in Modified Gravity: A Comparative Study of Approximations and $N$-Body Simulations}

\author[]{B. Bose$^{1,2}$\thanks{E-mail: \href{mailto:ben.bose@ed.ac.uk}{ben.bose@ed.ac.uk}}, 
A. Sen Gupta${^{3,4}}$
B. Fiorini${^{4}}$, 
G. Brando${^5}$, 
F. Hassani${^6}$, 
T. Baker${^4}$, 
\newauthor 
L. Lombriser${^7}$, 
B. Li${^8}$, 
C. Ruan$^{6}$,  
C. Hern\'andez-Aguayo$^{9,10}$, 
L. Atayde$^{11}$, 
N. Frusciante$^{12}$  \\ \\ 
$^{1}$Institute for Astronomy, University of Edinburgh, Royal Observatory, Blackford Hill, Edinburgh, EH9 3HJ, UK \\ 
$^{2}$Basic Research Community for Physics e.V., Mariannenstraße 89, Leipzig, Germany \\
$^{3}$Astronomy Unit, School of Physical and Chemical Sciences, Queen Mary University of London, Mile End Road, London, E1 4NS, UK \\ 
$^{4}$Institute of Cosmology and Gravitation, University of Portsmouth, Burnaby Road, Portsmouth PO1 3FX, UK \\ 
$^{5}$Max Planck Institute for Gravitational Physics (Albert Einstein Institute) Am M\"uhlenberg 1, 14476 Potsdam-Golm, Germany \\ 
$^{6}$Institute of Theoretical Astrophysics, University of Oslo, P.O. Box 1029 Blindern, 0315 Oslo, Norway \\
$^{7}$D\'epartement de Physique Th\'eorique, Universit\'e de Gen\`eve, 24 quai Ernest Ansermet, 1211 Gen\`eve 4, Switzerland \\
$^8$Institute for Computational Cosmology, Department of Physics,
Durham University, South Road, Durham, DH1 3LE, UK \\
$^{9}$Max-Planck-Institut f\"ur Astrophysik, Karl-Schwarzschild-Str. 1, D-85748, Garching, Germany\\%
$^{10}$Excellence Cluster ORIGINS, Boltzmannstrasse 2, D-85748 Garching, Germany\\%
$^{11}$Instituto de Astrofis\'ica e Ci\^{e}ncias do Espa\c{c}o, Faculdade de Ci\^{e}ncias da Universidade de Lisboa, Edificio C8, Campo Grande, P-1749016, Lisboa, Portugal\\
$^{12}$Dipartimento di Fisica ``E. Pancini", Universit\`a degli Studi  di Napoli  ``Federico II", Compl. Univ. di Monte S. Angelo, Edificio G, Via Cinthia, I-80126, Napoli, Italy\\
}
\date{Accepted XXX. Received YYY; in original form ZZZ}

\pubyear{2022}

\begin{document}
\label{firstpage}
\pagerange{\pageref{firstpage}--\pageref{lastpage}}
\maketitle



\begin{abstract}
Testing gravity and the concordance model of cosmology, $\Lambda$CDM, at large scales is a key goal of this decade's largest galaxy surveys. Here we present a comparative study of dark matter power spectrum predictions from different numerical codes in the context of three popular theories of gravity that induce scale-independent modifications to the linear growth of structure: nDGP, Cubic Galileon and K-mouflage. In particular, we compare the predictions from full $N$-body simulations, two $N$-body codes with approximate time integration schemes, a parametrised modified $N$-body implementation and the analytic halo model reaction approach. We find the modification to the $\Lambda$CDM spectrum is in $2\%$ agreement for $z\leq1$ and $k\leq 1~h/{\rm Mpc}$ over all gravitational models and codes, in accordance with many previous studies, indicating these modelling approaches are robust enough to be used in forthcoming survey analyses under appropriate scale cuts. We further make public the new code implementations presented, specifically the halo model reaction K-mouflage implementation and the relativistic Cubic Galileon implementation. 
\end{abstract}



\section{Introduction}
\label{sec:Introduction}

Observations of cosmological large-scale structure (LSS) offer a unique laboratory in which to test the concordance cosmological model, $\Lambda$CDM, which assumes General Relativity (GR). Such experiments are highly motivated. Indeed, the nature of the cold dark matter (CDM) and the constant dark energy ($\Lambda$) components, constituting 95\% of the Universe's total energy density \citep[see for example][]{SupernovaSearchTeam:1998fmf,SupernovaCosmologyProject:1998vns,eBOSS:2020yzd,Aghanim:2018eyx}, remains elusive. Moreover, $\Lambda$CDM's inability to reconcile principles of GR with quantum mechanics points to the need for a more unified theory~\citep[see][for a recent review on gravitational approaches to the cosmological constant problem]{Bernardo:2022cck}. These gaps in our understanding motivate the investigation into alternative theories beyond $\Lambda$CDM. By exploring these new frontiers, we hope to uncover a more comprehensive picture of the Universe, potentially leading to groundbreaking insights into its origin, evolution, and ultimate fate.

This decade will provide an immense opportunity for such insights through the efforts of some of the biggest scientific collaborations to date. These include the European Space Agency's Euclid mission \citep{Euclid:2024yrr}, the Vera Rubin Observatory \citep[][]{Abate:2012za,LSST:2008ijt} (LSST)\footnote{Vera Rubin was formerly known as the Large Synoptic Survey Telescope.}, the Dark Energy Survey \citep{Albrecht:2006um,DES:2016jjg}, the Nancy Grace Roman Space Telescope \citep{2019arXiv190205569A} and the Dark Energy Spectroscopic Instrument \citep{Levi:2019ggs}. For instance, Euclid and LSST will be measuring up to order 1 billion galaxy shapes \citep{Euclid:2024yrr,LSST:2008ijt}, 2 orders of magnitude more than previous surveys \citep[see for example][]{Hildebrandt:2016iqg}. This means the statistical precision of its resulting weak lensing measurements, such as cosmic shear, will be roughly the same order of magnitude better than previous observations, providing a potentially brilliant probe for new physics.

Consistency tests of $\Lambda$CDM are a primary goal, but these missions are also charged with investigating if there is any statistical preference for new physics. Such beyond-consistency tests require theoretical modelling of any new physics we wish to test. In particular, a key task is to theoretically model key statistical cosmological quantities over a very wide range of physical scales. The 2-point correlation function, or its Fourier analog, the power spectrum, of the cosmological matter distribution is one such summary statistic. At small physical scales, where we have many more galaxy pairs, the measured statistics will be far more precise, potentially providing a heightened signal of any new physics. It should be kept in mind that this work only considers the matter power spectrum, which is a key ingredient for cosmic shear weak lensing analyses.

The small scale precision measurements of forthcoming surveys has forced  ambitious accuracy demands on such theoretical predictions \citep[for example $\mathcal{O}(1\%)$ accuracy on the matter power spectrum][]{Hearin:2011bp,LSST:2008ijt,Euclid:2020tff}. This requires careful consideration of scale cuts. Most Euclid forecasts~\citep{Euclid:2019clj,Euclid:2022hdx,Euclid:2023rjj,Euclid:2023tqw} consider a `pessimistic' and `optimistic' scale cut in harmonic space, corresponding to a maximum angular multipole of $\ell=1500$ and $\ell=5000$, with the precise value of these cuts in Fourier mode, or $k$-space, varying with redshift. In contrast, LSST applies scale cuts in real space. While percent-level accuracy remains a desirable goal, definitive accuracy up to a specific $k$-cut is only achievable through full parameter inference. This nuanced approach is essential to fully harness the power of these surveys for an exquisite and reliable test of gravity and cosmology.

For these reasons, the community has sought to accurately model these small, nonlinear scales in the matter power spectrum, for beyond-$\Lambda$CDM scenarios. To this end, many methods have been developed to provide such predictions. $N$-body simulations provide our most accurate predictions, and have been extended to many models beyond-$\Lambda$CDM \citep[see for example][]{Li:2011vk,Li:2013nua,Li:2013tda,Puchwein:2013lza,Llinares:2013jza,Ruan:2021wup,Hernandez-Aguayo:2021kuh, Hassani:2019lmy, Hassani:2019wed, Christiansen:2023tfy}. This accuracy comes at a large computational cost, making this method inappropriate for expensive data-theory comparisons where we wish to sample a large cosmological and gravitational parameter space. One can alleviate this cost to some extent through approximate methods. For example, Comoving Lagrangian Acceleration (COLA) \citep{Tassev:2013pn,Howlett:2015hfa} is an $N$-body method that provides a balance between accuracy and speed by reducing the time steps in particle evolution through the perturbative modelling of large scale physics. This method has also been extended to many alternatives to $\Lambda$CDM~\citep{Winther:2017jof,Wright:2022krq,Brando:2023fzu}.

While being faster, COLA methods are still too slow to use directly in data analyses. Despite the computational cost, simulation methods are essential in bench-marking or constructing faster predictive pipelines, such as emulators \citep{Arnold:2021xtm,Harnois-Deraps:2022bie,Ramachandra:2020lue,Fiorini:2023fjl, Nouri-Zonoz:2024dph} or analytic models \citep{Zhao:2013dza,Mead:2016zqy}.  The halo model reaction \citep{Cataneo:2018cic} is one such analytic method, which can provide a high accuracy at a fraction of the time cost and is theoretically general, allowing its extension to many models of cosmology.

This paper is dedicated to assessing the consistency of these different methods for a few representative beyond-$\Lambda$CDM models of cosmological relevance. The models we consider are the DGP braneworld model~\citep{Dvali:2000hr}, the Cubic Galileon model~\citep{Nicolis:2008in} and the K-mouflage model~\citep{Babichev:2009ee}. This work runs in a similar vein to the code comparison project of Ref.~\cite{Winther:2015wla}, updating the exercise, nearly a decade later, to account for improvements in the codes and methods, as well as approximations and new theoretical models and phenomenology. Such an assessment is vital in modelling the theoretical uncertainty or delimiting the scales of validity of the method under consideration, which will play an important role in forthcoming surveys \citep{Audren:2012vy,Baldauf:2016sjb}. We also present an extension of the halo model reaction code, {\tt ReACT}, which includes the specific K-mouflage model of gravity considered in this paper. 

We outline the paper as follows: In \autoref{sec:theory} we briefly introduce the different beyond-$\Lambda$CDM models we consider. In \autoref{sec:tools} we outline the different methods we will compare, highlighting the key differences between them and the various approximations they employ. In \autoref{sec:results} we present matter power spectrum boost comparisons of the different methods. We present our conclusions in \autoref{sec:conclusions}.

\subsection{Notation and conventions}

In this work we will use the following definitions and conventions:
\begin{enumerate}
    \item 
    We use a metric signature of $(-,+,+,+)$. 
    \item 
    We work in units where $c=\hbar = 1$. 
    \item 
    Jordan frame quantities appear with a hat, e.g. $\hat{q}$.
    \item 
    The Planck mass is denoted as $M_{\rm pl}=(8\pi G_{\rm N})^{-1}$, where $G_{\rm N}$ is Newton's constant.
    \item 
    Overdots denote derivatives with respect to cosmic time $t$.
    \item 
    Primes denote derivatives with respect to the natural logarithm of the scale factor, $\ln a$, unless otherwise stated.
    \item 
    Quantities with a `$0$' subscript denote their value at $z=0$.
    \item 
    The canonical scalar field kinetic energy is $X \equiv -(\partial \phi)^2/2 $. 
\end{enumerate}


\section{Gravity Beyond General Relativity}
\label{sec:theory}

The simplest, viable class of alternatives to $\Lambda$CDM can be found by adding a single extra scalar degree of freedom, $\phi$, to GR. Under some basic constraints, such as second-order equations of motion (a generic condition to avoid unbounded negative energies) and four spacetime dimensions, the well-studied Horndeski Lagrangian encompasses all possible scalar-tensor theories with minimally coupled matter \citep{Horndeski:1974wa,Deffayet:2011gz,Kobayashi:2019hrl}. If we accept the speed of light to be the same as gravitational wave propagation,  in accordance with the observation of gamma ray burst GRB 170817A \citep{Goldstein:2017mmi}  and merger signal of GW170817 \citep{Monitor:2017mdv}, the Horndeski Lagrangian reduces to \footnote{This condition may not hold below the frequency band of terrestrial gravitational wave detectors \citep{deRham:2018red,deRham:2021fpu, LISACosmologyWorkingGroup:2022wjo, Baker:2022eiz, Harry:2022zey}.} \citep{Lombriser:2015sxa,Lombriser:2016yzn,Creminelli:2017sry,Ezquiaga:2017ekz,Baker:2017hug,Sakstein:2017xjx,Battye:2018ssx,deRham:2018red,Creminelli:2018xsv,Quartin:2023tpl}
\begin{align}
\mathcal{L}_{\rm H}  = G_4(\phi)\,R + G_2(\phi,X) - G_3(\phi,X)\Box\phi  \, , 
\label{eq:horndeski}
\end{align}
where $R$ is the Ricci curvature scalar, $\Box$ is the D'Alembert operator and each $G_i(\phi,X)$, $i=2,3,4$ is a free function of the scalar field $\phi$ and its canonical kinetic term $X$. Note that the $G_4$ operator is a function of $\phi$ only.

Besides modifying the expansion history of the Universe, modified gravity theories
 also leave an impact on the growth of structure~\citep[see][for a review]{Hou:2023kfp}. This is generally understood by considering linear perturbations on top of a homogeneous and isotropic Friedmann-Lemaître-Robertson-Walker background given by the following line element
\begin{equation}\label{eq:met_pert_flrw}
    \mathrm{d} s^{2} = - \left( 1 + 2\Psi \right)\mathrm{d} t^{2} + a^{2}(t)\left( 1 - 2 \Phi \right)\delta_{ij}\mathrm{d} x^{i} \mathrm{d} x^{j},
\end{equation}
where $\Phi$ is the usual Poisson potential in Newtonian gravity that captures perturbations in the spatial sector of the metric, while $\Psi$ is a gravitational potential corresponding to perturbations in the time-like sector of the line element.

The linear evolution of perturbations of modified gravity theories given by \autoref{eq:horndeski} has been thoroughly studied by many different works in the literature~\citep[see for example][]{Hu:2013twa,Zumalacarregui:2016pph,Frusciante:2019xia}. Within the quasi-static approximation~\citep{Sawicki:2015zya,Winther:2015pta,Pace:2020qpj}, the effects of modified gravity on the linear growth of structure in the Universe are encoded in a time- and scale-dependent effective gravitational constant 
\begin{equation}\label{eq:lineargeff}
    G_{\rm eff, L}(k,a) = G_{\rm N}\left[ 1 + \frac{\Delta G_{\rm eff, L}(k,a)}{G_N}\right],
\end{equation}
where $k$ is the Fourier mode. In this work, we only consider theories where the linear modification is scale-independent and so we drop the dependency on $k$ for $G_{\rm eff, L}$, where L refers to a linear theory prediction. The Poisson equation at large scales is then written in Fourier space as 
\be\label{eq:poi_mg}
k^2 \Phi(k,a)=4 \pi G_{\rm eff, L}(a) a^2 \bar{\rho}_{\mathrm{m}} (a) \delta_{\rm m}(k,a),
\ee 
where $\bar{\rho}_{\rm m}$ is the background matter density, and $\delta_{\rm m}$ is the corresponding linear matter perturbation.

Another requirement for this class of theories is the inclusion of a theoretical mechanism that prevents large modifications in environments where GR-like physics has been well confirmed by experiment \citep[see][for example]{Will:2014kxa,Belgacem:2018wtb}. Such mechanisms are known as screening mechanisms \citep[see][for a recent review and experimental tests]{Brax:2021wcv}. The screened environments are thus  small scale, dense environments. This means that the modification to Newton's constant, more generally written as 
\begin{equation}\label{eq:nlgeff}
    G_{\rm eff}(k,a) = G_{\rm N}\left[ 1 + \frac{\Delta G_{\rm eff}(k,a)}{G_N}\right],
\end{equation}
requires the condition that $\lim_{k\to \infty} G_{\rm eff}(k,a)\to G_{\rm N}$. In this case $G_{\rm eff}(k,a)$ is the effective gravitational constant valid at all scales - both linear and nonlinear - and it necessarily depends on scale as well as time. In this work we will meet two such screening mechanisms which satisfy this condition: the Vainshtein mechanism and K-mouflage screening.

Returning to \autoref{eq:horndeski}, we will consider three choices for the Lagrangian functions, each having very particular phenomenological features, including different screening mechanisms and cosmological backgrounds. Where a choice exists, we will give their Lagrangians in the Einstein frame where $G_4(\phi) = M_{\rm pl}^2/2$, with metric $g_{\mu \nu}$. In this frame the `pure gravity' part of the action resembles the Einstein-Hilbert action for GR, simplifying some computations. However, this frame choice also results in non-minimal coupling of matter to the metric, ensuring the theory behaves very differently to GR.

The Einstein frame is obtained by performing a conformal transformation of the Jordan frame. The Jordan frame prioritises use of a metric, $\hat{g}_{\mu \nu}$, which couples minimally to the matter fields but contains the non-trivial $G_4$ function. In this frame the gravitational Lagrangian is modified from the the Einstein-Hilbert action. The Jordan-frame metric is related to the Einstein-frame metric, $g_{\mu \nu}$, via a conformal factor $A$ that is a function of the Horndeski scalar: 
\begin{equation}\label{eq:conformalfacdef}
 \hat{g}_{\mu \nu} = A^2(\phi) g_{\mu \nu} \, .     
\end{equation} 
In what follows, specifically in the case of K-mouflage theories, we will see that some quantities differ between the Jordan and Einstein frame. Though these quantities may be `physical' in nature, they are not directly observable. General coordinate invariance -- a key property shared with GR by nearly all modified gravity theories -- ensures that observable quantities must be independent of frame choices \citep[see for example][]{Catena:2006bd,Chiba:2013mha,Francfort:2019ynz}.
\\
\\
We summarise the models considered in this paper, their associated additional parameters and some selected constraints in \autoref{tab:models}. 
\begin{table*}
    \centering
    \caption{Overview of gravity models considered in this work. Note the K-mouflage kinetic term in \autoref{eq:kmnckinetic} does not pass Solar System tests without running into fine-tuning issues \citep{Barreira:2015aea}. Note the CG has no free parameters with the tracker solution. We have included constraints for the more general GCCG (see \autoref{sec:cubicgalileon}).}
    \label{tab:models}
    \bigskip
    \begin{tabular}{l||l|l|l}
       model & screening method & free parameters & selected data constraints    \\
 \hline
nDGP & Vainshtein & $ \{ \Omega_{\rm rc} \} $  & $\Omega_{\rm rc} \leq 0.235 \, (2 \sigma)$   (LSS) \citep{Barreira:2016ovx}  \\ 
CG & Vainshtein &  $\{s=2,q=0.5\}$ & $s = 0.05^{+0.08}_{-0.05}$, $q>0.8$ $(2 \sigma)$ (Various LSS, GCCG) \citep{Frusciante:2019puu}  \\ 
K-mouflage & K-mouflage & $\{n,\beta_{\rm K}, K_0, \lambda \} $ &$\beta_{\rm K} \leq 0.1$  (Lunar laser ranging) \citep{Barreira:2015aea}  \\ 
    \end{tabular}
\end{table*}


\subsection{nDGP}\label{sec:dgp}
The first model we consider is the  Dvali-Gabadadze-Porrati model~\citep{Dvali:2000hr}, which does not strictly fall into the Horndeski class, being a five-dimensional braneworld model. It is given by the following action
\begin{equation}\label{eq:dgp5d}
S = \frac{1}{16\pi G_5}\int_{\cal M} {\rm d}^5 x \sqrt{-\gamma} R_5
+ \int_{\partial {\cal M}} {\rm d}^4 x \sqrt{-g} 
\left[\frac{M_{\rm pl}^2}{2} R + {\cal L}_{\rm m} \right]\,,
\end{equation}
where $\gamma$ is the five-dimensional metric and $R_5$ its Ricci curvature scalar. $G_5$ is the five-dimensional gravitational constant. 
The matter Lagrangian is restricted to a four-dimensional brane in a five-dimensional Minkowski spacetime. The induced gravity given by the four-dimensional Einstein-Hilbert action is responsible for the recovery of four-dimensional gravity on the brane. The parameter $r_{\rm c} = G_5/(2 G_{\rm N})$ is called the cross-over scale and is the only free parameter of the model, with its GR limit being $r_{\rm c} \to \infty$. 

DGP also exhibits screening coming from higher order derivative terms in the effective 4-dimensional action. Such screening is known as Vainshtein screening \citep{Vainshtein:1972sx,Babichev:2013usa}. The so-called decoupling limit of DGP has the effective action given by \citep{Luty:2003vm,Gabadadze:2006tf} 
\begin{equation}\label{eq:dgpdecouple}
    \mathcal{L}_{\rm DGP} = \frac{M_{\rm pl}^2}{2} R + \left( 3 \phi - \frac{1}{\Lambda^3_{\rm DGP}} (\partial \phi)^2 \right) \Box \phi + \frac{1}{2 M_{\rm pl}^2} \phi T \, , 
\end{equation}
where $T$ is the trace of the energy momentum tensor and $\Lambda^3_{\rm DGP} = M_{\rm pl}^2/r_c^2$. Note that although \autoref{eq:dgp5d} is not a Horndeski Lagrangian, \autoref{eq:dgpdecouple} is (compare to \autoref{eq:horndeski}). In this case we have 
\begin{equation}
    G_3(\phi, X) = 3 \phi + \frac{2}{\Lambda_{\rm DGP}^3} X \, , 
\end{equation}
with $G_2(\phi, X) = 0 $ and $G_4(\phi) = M_{\rm pl}^2/2$.

The literature typically assumes a $\Lambda$CDM  background expansion, which is accommodated by introducing an appropriate dark energy contribution \citep[see for example][]{Schmidt:2009sv,Bag:2018jle} on the stable `normal' branch solution of the Friedmann equations. We follow this here \citep[see][for more details]{Lue:2005ya} and refer to this normal branch as nDGP. We also parametrize the modification to gravity using the energy density fraction $\Omega_{\rm rc} \equiv 1/(4r_{\rm c}^2H_0^2)$, where $H_0$ is the Hubble constant. The GR-limit is then $\Omega_{\rm rc} \rightarrow 0$. 

Although nDGP is now quite strongly constrained by observations~\citep[see for example][]{Lombriser:2009xg,Barreira:2016ovx,Piga:2022mge}, its appeal as a modified gravity model stems from the simplicity of its 4D effective action relative to the new phenomenology it introduces. It is one of the simplest examples of a gravity model that produces Vainshtein screening effects, whilst maintaining scale-independent growth of matter perturbations, and having only one additional parameter relative to $\Lambda$CDM. This has made it a favourite testbed for simulations \citep{Schmidt:2009sg,Khoury:2009tk,Li:2013nua,Winther:2017jof} and analyses with galaxy surveys \citep{Barreira:2016ovx,Piga:2022mge,Euclid:2023rjj}. We refer the reader to Refs.~\cite{Koyama:2005kd,Li:2013nua} and \autoref{sec:ppfgeff} for details on the modification to the Poisson equation (\autoref{eq:poi_mg}) in linear and nonlinear regimes.


\subsection{Cubic Galileon} \label{sec:cubicgalileon}
The Cubic Galileon (CG) model was first derived by~\cite{Nicolis:2008in} as a generalisation of the effective DGP action in 4-dimensions. The Lagrangian is given by \citep[see for example][]{Deffayet:2009wt,Kobayashi:2010cm}
\begin{equation}\label{eq:cG}
    \mathcal{L}_{\rm CG} = R\frac{M_{\rm pl}^2}{2} + c_2 X + \frac{1}{\Lambda_3^3} c_3 X \Box \phi \, , 
\end{equation}
where $c_2$ and $c_3$ are dimensionless constants parametrizing the modification to gravity, and the canonical choice for $\Lambda_3$ being $\Lambda_3^3 = M_{\rm pl}H_0^2$, made to give the scalar field non-trivial dynamics on cosmological scales. Comparing with \autoref{eq:horndeski} we have
\begin{equation}
G_2(\phi,X) = c_2 X \, , \qquad G_3(\phi,X) = - \frac{1}{M_{\rm pl} H_0^2} c_3 X \, . 
\label{eq:covariantgal}
\end{equation}
This model also exhibits the Vainshtein mechanism due to the presence of the higher-order derivative terms \citep[see][for a derivation in the case of spherical symmetry]{Barreira:2013eea}. In this model, $G_4(\phi) = A(\phi)^{-2}/2 = 1$, and hence there is no difference between Jordan and Einstein frames (see \autoref{eq:conformalfacdef}). We note that the absence of $G_4$ and conformal coupling allows one to interpret this model as a dark energy model with a non-trivial kinetic term.

The Cubic Galileon model is one member of a broader family, the Galileons, which added further derivative terms to \autoref{eq:cG} \citep{Deffayet:2009wt}. The Galileon family received intense interest from the theoretical physics community due to their shift symmetry properties (the actions are invariant under a shift $\phi(x) \rightarrow \phi(x) +c+b_{\mu} x^\mu$, $c$ and $b_\mu$ constants); this leads to special properties of the S-matrix. Cosmologically, their impact has been studied on the CMB \citep[for example][]{Barreira:2014jha,Peirone:2019aua,Frusciante:2019puu,Albuquerque:2021grl}, linear matter power spectrum \citep[for example][]{Barreira:2012kk} and gravitational lensing by voids \citep[for example][]{Baker:2018mnu}. See also Refs.~\cite{Renk:2017rzu,Peirone:2017vcq,Frusciante:2020zfs} for other observational implications.

The more complex Galileon siblings have been virtually eliminated by their inability to have luminal gravitational waves, leaving behind only the Cubic Galileon \citep[see for example][]{Ezquiaga:2017ekz, Baker:2017hug}. The Cubic Galileon model can be constrained by considering the integrated Sachs-Wolfe effect cross-correlated with a galaxy sample, as was done in Refs.~\cite{Renk:2017rzu,Kable:2021yws}. The resulting cross-correlation, however, is shown to be anti-correlated with the expected $\Lambda$CDM signal, which severely constrains this model. It is worth noting, nevertheless, that a broader class of cubic Horndeski theories does not show this anti-correlation~\citep{Brando:2019xbv}. Similarly to nDGP, it remains a useful testbed displaying Vainshtein screening, with a larger degree of flexbility due to its additional parameters and energy scales. It is a useful starting point from which to investigate the more general model space of the Horndeski Lagrangian (\autoref{eq:horndeski}).

We also note that the non-zero $G_2$ term makes this model phenomenologically distinct from nDGP. Further, in this paper  we do not assume a $\Lambda$CDM background as with nDGP, but rather the solution to the Friedmann equations which include the effects of the scalar field \citep[see for example][]{Barreira:2013eea}. A cosmology with this background evolution but with no further gravitational modification (so the Poisson equation remains as in GR), will  be referred to as QCDM as in Ref.~\cite{Barreira:2013eea}.

The more general Generalised Covariant Cubic Galileon (GCCG) was recently considered in  Ref.~\cite{Frusciante:2019puu}, which promotes the $G_i$ functions to be power law functions of $X$, i.e. $G_i \propto X^{p_i}$. This model permits a tracker solution at the background level which is given by~\citep{DeFelice:2011bh}
\begin{equation}
    H^{2q +1} \psi^{2 q}  = \zeta H_0^{2q +1} \, ,
\end{equation}
where $q \equiv (p_3 - p_2) + 1/2$ and $\psi = \phi' /M_{\rm pl}$. We also have the parameter $s = p_2/q$, leaving only 2 additional degrees of freedom for this model over $\Lambda$CDM. The GCCG reverts to the Cubic Galileon model when $q = 0.5$ and $s=2$. 

The GCCG model has not been ruled out by data, with CMB experiments giving the $2 \sigma$ bounds of $q>0$ and $s = 0.6^{+1.7}_{-0.6}$, with a slight preference for the model over $\Lambda$CDM~\citep{Frusciante:2019puu}. When combined with SN1a and redshift space distortion data sets, the bounds improve to $q>0.8$ and $s = 0.05^{+0.08}_{-0.05}$. We note that theoretical stability conditions require both parameters to be positive.
\\
\\
In this paper we will only consider the CG limit of GCCG. We note that we employ the GCCG patch to the {\tt ReACT} code~\citep{Atayde:2024tnr} for those specific predictions. For details on how the Poisson equation is modified in the CG limit, we refer the reader to Refs.~\cite{Barreira:2013eea,Atayde:2024tnr}.


\subsection{K-mouflage} \label{sec:kmouflage}
\subsubsection{Lagrangian}

The last model we consider is the K-mouflage model \citep{Babichev:2009ee}. This model has the Lagrangian (in the Einstein frame)
\begin{equation}
    \mathcal{L}_{\rm K} =  R \frac{M_{\rm pl}^2}{2} + \mathcal{M}^4 K(X) \, , 
\end{equation}
where $K(X)$ is a function of the canonical kinetic term, equivalent to a restricted $G_2(\phi, X)$, and $\mathcal{M}^4$ is an energy scale of the theory\footnote{Not to be confused with the manifold $\mathcal{M}$ in \autoref{eq:dgp5d}.}. We will set $\mathcal{M}^4 = \lambda^2 H_0^2 M_{\rm pl}^2$ as in Ref.~\cite{Hernandez-Aguayo:2021kuh}, $\lambda$ being an order $1$ dimensionless constant which can be tuned to give the current accelerated expansion of the Universe today. In this work we will consider a form which has been well studied in the literature \citep{Brax:2014wla,Brax:2014yla,Barreira:2014gwa,Barreira:2015aea,Hernandez-Aguayo:2021kuh} 
\begin{equation}
K(X) = -1 + \frac{1}{H_0^2 \lambda^2 M_{\rm pl}^2 } X + K_0 \frac{1}{H_0^{2n} \lambda^{2n} M_{\rm pl}^{2n} } X^n  \, ,  \label{eq:kmnckinetic}
\end{equation}
where $K_0$ is another dimensionless model parameter and $n \geq 2$ is an integer. For the conformal function, we assume an exponential form
\begin{equation}
    A(\phi) = \exp\left(\frac{\beta_{\rm K} \phi}{M_{\rm pl}}\right) \, , 
    \label{eq:conffac}
\end{equation}
where $\beta_{\rm K}$ is another dimensionless model parameter. In total we then have 4 parameters for this particular model: $\{\lambda, K_0, n, \beta_{\rm K} \}$. 

Unlike the other two models considered, the Jordan and Einstein frames are not set to be identical ($A(\phi)\neq1$) which distinguishes this model from $k$-essence theories \citep{Armendariz-Picon:1999hyi} where a universal coupling to matter is not present. In this work we will develop predictions for both frames. We provide the transformations of key quantities in the next subsection.

This model exhibits a similar screening mechanism to Vainshtein screening, although quantitatively different due to the absence of the higher-order $G_3(\phi, X) \Box \phi$ term, giving it a unique phenomenology. In particular, in dense environments of mass $m$, the K-mouflage radius -- the scale below which GR is recovered -- goes as $m^{1/2}$, whereas in Vainshtein theories this screening occurs at smaller scales, with a dependence of the Vainshtein radius on the environmental mass being $m^{1/3}$~\citep{Brax:2014wla}. Vainshtein is also capable of screening large cosmological structures, while K-mouflage is not \citep{Brax:2015lra}. 

K-mouflage has been confronted with a number of cosmological data sets in Refs.~\cite{Barreira:2015aea,Benevento:2018xcu}, with a review of current constraints given in Ref.~\cite{Brax:2021wcv} and forecasts using spectroscopic and photometric primary probes by Euclid given in  Ref.~\cite{Euclid:2023rjj}. In particular, in Ref.~\cite{Barreira:2015aea}, the authors place a Solar System constraint on the coupling parameter $\beta_{\rm K} \leq 0.1$, and argue that the power law form for $K(X)$ as chosen here will necessarily require a degree of fine tuning to avoid constraints. Despite this, this model is a good test case for implementation as it has been well studied in the literature and there are available $N$-body simulations with which to compare to \citep{Hernandez-Aguayo:2021kuh}. More viable non-canonical kinetic terms can easily be implemented following the current implementations.

We alert the reader that we have made public a \href{https://github.com/nebblu/ACTio-ReACTio/tree/master/notebooks}{\texttt{Mathematica} notebook}
with some key Einstein frame quantities and derivations for the model along with this work. This contains useful expressions such as the exact solutions for the Einstein frame background $H(a)$ in the $n=2$ and $n=3$ cases.


\subsubsection{Transformation to Jordan Frame} \label{sec:frames}
In this section we provide some basic translations between Einstein and Jordan frames which will be useful for our comparisons of the K-mouflage model. We follow Ref.~\cite{Francfort:2019ynz} for these expressions. We use subscripts `J' and `E' to denote Jordan and Einstein frame quantities respectively. 

The scale factor transforms as 
\begin{equation}
    a_{\rm J} = \bar{A} \, a_{\rm E} \, ,
    \label{eq:scalefacframe}
\end{equation}
where $\bar{A}$ is the conformal factor evaluated at the background level (see \autoref{eq:conffac}). The Hubble rate transforms as  
\begin{equation}
    H_{\rm J}(a) = \frac{H_{\rm E} }{\bar{A}} \left[ 1 + \frac{\beta_{\rm K}}{M_{\rm pl}} \frac{{\rm d} \phi}{{\rm d } \ln a_{\rm E}}  \right] \, . 
\end{equation}
The matter power spectrum transforms as \citep{Francfort:2019ynz} 
\begin{align}
  (2 \pi)^3 \delta_{\rm D}(\bfk_1 + \bfk_2) P_{\rm J}(k_1) & = \langle \delta_{\rm J} (\bfk_1) \delta_{\rm J}(\bfk_2) \rangle \nonumber \\ 
  & = \langle \delta_{\rm E}(\bfk_1) \delta_{\rm E}(\bfk_2)\rangle \nonumber \\ 
  & - 4 \frac{\bar{A}_\phi}{\bar{A}} \langle \delta_{\rm E}(\bfk_1) \delta \phi(\bfk_2)\rangle \nonumber \\ & - 4 \frac{\bar{A}_\phi}{\bar{A}} \langle \delta \phi(\bfk_1) \delta_{\rm E}(\bfk_2)\rangle \nonumber \\ & + 16 \left(\frac{\bar{A}_\phi}{\bar{A}}\right)^2 \langle \delta \phi(\bfk_1) \delta \phi(\bfk_2)\rangle \, , \label{eq:pktrans}
\end{align}
where and we used $\delta_{\rm J} = \delta_{\rm E} - 4 \delta \phi \bar{A}_\phi/\bar{A}$, with $\bar{A}_\phi = {\rm d} \bar{A}(\phi)/{\rm d}\phi$, $\delta$ is shorthand for the matter density field perturbation $\delta_{\rm m}$, $\delta \phi$ is the scalar field perturbation, $\phi = \bar{\phi} + \delta \phi$, and $k$ is the comoving Fourier mode in $h/{\rm Mpc}$. Angular brackets denote an ensemble average. The linear order Klein-Gordan equation for the scalar field perturbation in Fourier space under the quasi-static approximation is \citep{Brax:2014yla} 
\begin{equation}
\delta \phi (\bfk) = -\frac{\bar{A} \beta_{\rm K} a^2 }{M_{\rm pl} K_X k^2} \bar{\rho}_{\rm m} \delta_{\rm E}(\bfk) \,,  
\end{equation}
 where $K_X = {\rm d} \, K(X)/ {\rm d}X$. Substituting $\delta \phi$ into \autoref{eq:pktrans} gives us the following relationship between linear matter power spectra predictions
\begin{equation}
P_{\rm L,J}(k) =  P_{\rm L,E}(k) \left[ 1 + 2 \mathcal{J}(k,a) +  \mathcal{J}(k,a)^2 \right] \, , 
\end{equation}
where
\begin{equation}
    \mathcal{J}(k,a) = \frac{12 \bar{A}_\phi \beta_{\rm K} H_0^2 M_{\rm pl} \Omega_{\rm m,0}}{ a k^2 K_X} \, ,
\end{equation}
where we have used the relation $\bar \rho_{\rm m} = 3 H_0^2 M_{\rm pl}^2 \Omega_{\rm m,0} a^{-3}$. We see that the linear matter power spectra in both frames are identical up to corrections that are suppressed by powers of $\sim H_0^2/k^2$.

It was argued in Ref.~\cite{Francfort:2019ynz} that this correction to the matter power spectrum at nonlinear scales continues to go as $\sim H^4/k^4$, and so becomes negligible on all sub-horizon scales. This argument hinged on a number of assumptions, including $A_\phi \sim - A(\phi)/(2 \phi)$. We will show in \autoref{sec:results} that the corrections are indeed small at nonlinear scales for the K-mouflage model, using a conformal factor given in \autoref{eq:conffac}.


\section{Tools \& Methods}
\label{sec:tools}

In this section we give an overview of the methods developed to give predictions for the large-scale structure in all modified gravity scenarios considered. After explaining details of how we compute matter power spectra, we describe the methods we will compare in this work. Most of these are $N$-body simulation-based approaches with various degrees of approximation. The halo model reaction \citep{Cataneo:2018cic} is also considered, which is an analytic method based on the halo model and perturbation theory. \autoref{tab:tools} gives an overview of these methods.

\begin{table*}
    \centering
    \caption{Overview of the numerical codes employed in this comparison. For more information on screening approximations see main text. PPF: parametrised post-Friedmannian; PM: particle mesh; AMR: adaptive mesh refinement; 2LPT: 2nd order Lagrangian perturbation theory; K-G: Klein-Gordon.}
    \label{tab:tools}
    \bigskip
    \begin{tabular}{l||l|l|l}
       code & type & screening approximation & reference(s) \\
 \hline
       {\tt ECOSMOG} & $N$-body (AMR) & full K-G solution & \cite{Li:2011vk} \\
       {\tt MG-GLAM} & $N$-body  (uniform PM) & full  K-G solution & \cite{Hernandez-Aguayo:2021kuh} \\
       {\tt MG-evolution} & $N$-body (uniform PM) & PPF with free parameter $k_*$  & \cite{Hassani:2020rxd,Adamek:2015eda,Adamek:2016zes} \\
       {\tt Hi-COLA} & $N$-body (PM in the 2LPT frame) & screening factor & \cite{Wright:2022krq}, Sen Gupta et al. (in prep.) \\
        {\tt COLA-FML} & $N$-body (PM in the 2LPT frame) & linear K-G equation in Fourier space & \cite{Winther:2017jof,Brando:2023fzu,Scoccimarro:2009eu} \\
       {\tt ReACT} & Halo model and perturbation theory& spherical collapse & \cite{Bose:2022vwi,Atayde:2024tnr} \\
    \end{tabular}
\end{table*}

\subsection{$P(k)$ estimation} \label{sec:estimation}
$N$-body simulations track the time-evolution of the matter distribution in the simulation box (of side $L_{\rm box}$) by means of a number of $N$-body particles ($N_{\rm P}$). To estimate the matter power spectrum from these sort of discrete distributions it is necessary to deal with some subtleties. The number of particles used in $N$-body simulations is often large (i.e. $10^{8}-10^{12}$) so that it would be computationally impractical to estimate the  matter power spectrum by computing the distances between each pair of particles. Hence, the particles are normally interpolated on a regular grid using mass assignment schemes (MAS). Then the  matter power spectrum is estimated exploiting the Fast Fourier Transform (FFT) algorithm. However, the modes close to the Nyquist frequency of the FFT grid can be significantly affected by aliasing \citep{Jing:2004fq,Sefusatti:2015aex}. To avoid this problem we use the interlacing technique with the triangular-shaped-cloud MAS \citep{Sefusatti:2015aex} to compute the matter power spectra from the simulations. Aiming to compare our matter power spectra deep in the nonlinear regime but mindful of the limited mass-resolution of our simulations, we use a FFT grid of size $N_{\rm mesh, 1D}= L_{\rm box}/({\rm d}x)$ where ${\rm d}x$ is the domain grid resolution of the simulation, 
and use a simple linear binning with $k_{\rm min} = k_{\rm f}/2$ and $\Delta k = k_{\rm f}$, where $k_{\rm f}\equiv \frac{2 \pi}{L_{\rm box}}$ is the fundamental frequency of the box.


\subsection{Full $N$-body} \label{sec:nbody}
Our reference predictions will come from numerical simulations that solve the nonlinear Klein-Gordon equation with multi-grid relaxation to get the precise modified force law. They also employ a large number of time steps over which the particles are evolved, ensuring the accuracy of the resulting predictions. We consider two variants of these codes.

\subsubsection{ECOSMOG} \label{sec:ecosmog}

The {\tt ECOSMOG} simulation code \citep{Li:2011vk,Li:2013nua} is a modified gravity extension of the adaptive mesh refinement (AMR) code Ramses \citep{Teyssier:2001cp}. {\tt ECOSMOG} relies on multigrid relaxation techniques to solve the nonlinear Klein-Gordon equations for the additional scalar fields that appear in some modified gravity theories (such as those considered here). This code was used to simulate several gravity models in the literature:
\begin{itemize}
    \item f(R) \citep{Li:2011vk};
    \item nDGP \citep{Li:2013nua}; 
    \item symmetron \citep{Davis:2011pj,Brax:2013mua};
    \item dilaton \citep{Brax:2011ja};
    \item galileon (cubic, quartic, cubic vector) \citep{Barreira:2013xea,Barreira:2013eea,Becker:2020azq}.
\end{itemize}
The accuracy of this code for predictions of $f(R)$~\citep{Hu:2007nk} effects on the matter power spectrum has been estimated to be of $\sim 1\%$ up to $k\sim 7 h/{\rm Mpc}$ in the code comparison paper Ref.~\cite{Winther:2015wla}. This code constitutes the highest precision predictive tool to be considered in this work.

\subsubsection{MG-GLAM} \label{sec:mgglam}
\texttt{MG-GLAM} \citep{Hernandez-Aguayo:2021kuh,Ruan:2021wup} extends the Particle Mesh (PM) code \texttt{GLAM} \citep{Klypin:2018MNRAS.478.4602K.GLAM} to a general class of modified gravity theories (including the K-mouflage model) by adding extra modules for solving the Klein-Gordon equations, using the multigrid relaxation algorithm. It uses a regularly spaced 3D mesh covering the cubic simulation box, solves the Poisson equation for the Newtonian potential using the Fast Fourier Transform (FFT) algorithm, and adopts the Cloud-In-Cell (CIC) scheme to implement the matter density assignment and force interpolation.

\texttt{MG-GLAM} has been tested with the results from other high-precision modified gravity codes, such as \texttt{ECOSMOG} \citep{Li:2011vk,Li:2013nua}, \texttt{MG-GADGET} \citep{Puchwein:2013lza}, and \texttt{MG-AREPO} \citep{Arnold:2019NatAs...3..945A,Hernandez-Aguayo:2021MNRAS.503.3867H}.
For example, using $1024^3$ particles in a box of size $512\, \mathrm{Mpc}/h$, \texttt{MG-GLAM} simulations can accurately predict the matter power boost, $P_{\mathrm{MG}} / P_{\mathrm{\Lambda CDM}}$ at $k \lesssim 3\,h / \mathrm{Mpc}$, with about $1\%$ of the computational costs of the high-fidelity code \texttt{ECOSMOG}. Being the only code that has been used in the literature to simulate K-mouflage cosmologies, an estimate of its accuracy for the K-mouflage boost factor is not available. However it has been compared to the tree-PM code \texttt{MG-Arepo} for another derivative coupling model (nDGP) where it showed an agreement of $\sim 2\%$ up to $k = 3\,h / \mathrm{Mpc}$, with deviations of $\sim 1\%$ from \texttt{MG-Arepo} (and theory predictions) already present on linear scales.

\subsection{MG-evolution} \label{sec:mgevo}


We further consider the relativistic $N$-body code, {\tt MG-evolution}~\citep[\href{https://github.com/FarbodHassani/MG-evolution}{\faicon{github}}][]{Hassani:2020rxd}. This code is based on {\tt gevolution} \citep{Adamek:2015eda}, integrating parametrised modifications of gravity for various dark energy scenarios. {\tt MG-evolution} is implemented based on a parametrisation framework that includes both linear and deeply nonlinear scales, where the nonlinear parametrisation is based on modified spherical collapse computations and a parametrised post-Friedmannian expression.

{\tt MG-evolution}  has been tested for a number of well-studied modified gravity models encompassing $f(R)$ and nDGP gravity that include large-field value and derivative screening effects \citep{Hassani:2020rxd}.  Unlike most modified gravity $N$-body implementations, {\tt MG-evolution} is as fast as the $\Lambda$CDM simulations as it does not need to deal with solving computationally expensive scalar field equations.

In \autoref{sec:ppfgeff} we discuss the nDGP and CG implementations in {\tt MG-evolution} through a parametrisation  with one screening transition, $k_*$, which is treated as a free parameter (see \autoref{sec:ppfgeff}). The effective gravitational constant is expressed as
\be
\frac{\Delta{G}_{\text {eff }}(k,a)}{G_{\rm N}}=\frac{\Delta G_{\text {eff, L}}(a)}{G_{\rm N}}  \times \frac{\Delta G_{\text {eff, NL}} (k,a)}{G_{\rm N}} \label{eq:param_mg_evolution}\, , 
\ee
where we recall $G_{\rm eff}(k,a)= G_{\rm N}[1+\Delta G_{\rm eff}(k,a)/G_{\rm N}]$, $G_{\text {eff, L}}$ denoting the linear regime parametrisation and $G_{\text {eff, NL}}$ refers to the  parametrisation of the nonlinear regime that includes the screening or other suppression effects. The expressions for $G_{\text {eff, NL}}$ are given in \autoref{sec:ppfgeff}. {\tt MG-evolution} then solves the modified Poisson equation (\autoref{eq:poi_mg}) based on $G_{\rm eff}$ obtained from \autoref{eq:param_mg_evolution}. It is worth noting that this parametrization of gravitational modification is done in Fourier space. As detailed in Ref.~\cite{Hassani:2020rxd}, this transformation yields an effective screening wavenumber $k_*$, which can be modeled \citep{Lombriser:2016zfz} for different screening types. Currently, as mentioned, we treat $k_*$ as a free parameter to be set by the user. In this work we tune the values of $k_*$ in order to optimize the agreement with the reference predictions in each model and at each redshift considered. The resulting values of $k_*$ are presented in \autoref{sec:results}.


\subsection{COmoving Lagrangian Acceleration} \label{sec:cola}

The COmoving Lagrangian Acceleration (COLA) method~\citep{Tassev:2013pn} is a hybrid $N$-body approach to performing dark matter simulations to study the effects of gravity on the formation of large-scale structure. It leverages the fact that the growth of structure on large scales can be computed analytically through Lagrangian Perturbation Theory (LPT). This informs the small-scale $N$-body part of COLA codes, thereby allowing for a significant speedup in the production of results at the cost of a modest loss of accuracy at small scales. In short, the COLA approach is a method well-suited for producing large-scale structure results on mildly nonlinear scales much faster than traditional $N$-body codes. 

The COLA method introduced by~\cite{Tassev:2013pn} was originally developed for use with GR $N$-body codes. However, being able to probe nonlinear scales is particularly useful in the study of modified gravity theories, as key phenomenology, such as the effects of screening mechanisms, become apparent on these scales. Since Tassev et al., implementations of COLA codes for modified gravity theories have followed for specific theories, such as $f($R) and nDGP~\citep{Winther:2017jof,Valogiannis:2016ane}. Below we describe two branches of work that extend the COLA method to more general families of gravity models.


\subsubsection{Hi-COLA} \label{sec:hicola}

Horndeski-in-COLA ({\tt Hi-COLA}) \citep[\href{https://github.com/Hi-COLACode/Hi-COLA}{\faicon{github}}][]{Wright:2022krq} is an implementation of the COLA methodology for a broad Horndeski class of scalar-tensor theories. {\tt Hi-COLA} aims not to carry hard-coded theory-specific implementations, but instead receives as input the Lagrangian functions for a given theory of interest, making it generic. The action of the new scalar degree of freedom, $\phi$, is included as a fifth force in the COLA simulation. The class of theories {\tt Hi-COLA} is designed to work with are \textit{reduced}-Horndeski theories, where gravitational waves travel at the speed of light. Such theories are described by \autoref{eq:horndeski}. This restriction follows the constraints on the theory space suggested by the analysis of GW170817\footnote{Though it should be noted that \autoref{eq:horndeski} is not the most general action describing theories that do not violate the results of GW170817. Some Gauss-Bonet theories are excluded, for example; see Ref.~\cite{Clifton:2020xhc}.}. 

After receiving inputs for the forms of the Horndeski functions, $G_2$, $G_3$ and $G_4$, the symbolic manipulation modules of {\tt Hi-COLA} construct the appropriate background equations of motion and background-dependent fifth force expressions and solves them. These are used to handle the expansion of the simulation box, compute 2nd-order LPT factors (2LPT) and construct the total force experienced by dark matter particles. This force can be schematically written as
\begin{align}
\label{eq:hicolaforce}
    F_{\textrm{total}} = G_{\rm eff} F_{\rm N} \,,
\end{align}
where
\begin{equation}
    G_{\textrm{eff}} = \frac{G_{\rm G4}}{G_{\rm N}} \big\{ 1 + \beta_{\rm HC}(z) S_{\textrm{HC}}(z,\delta_m)\big\}.
\end{equation}
%
$F_{\rm N}$ is the regular Newtonian force which is present in GR, and the multiplicative factor in braces represents the extra force contributions from $\phi$. $G_{G4}$ is the effective gravitational constant, which can differ from $G_{\rm N}$ in a time-dependent manner if $G_4$ in \autoref{eq:horndeski} is non-trivial. This term will play a role in the results of \autoref{sec:kmouflage_results}. 

\betahc \ is a background-dependent function known as the {\it coupling factor}; it controls the total possible strength of the fifth force at a given point in time. \shc \ is a background and density-dependent function called the \textit{screening factor}. On linear scales  $S_{\textrm{HC}} \rightarrow 1$, whilst in screened regimes  $S_{\textrm{HC}} \rightarrow 0$. Hence this factor is responsible for the suppression of the fifth force in on small scales, returning the theory's behaviour to GR. 

\shc \ is derived under a quasi-linear perturbative treatment, where the metric perturbations are considered to first order, whilst the scalar field \textit{derivative} perturbations are kept up to third order, following Ref.~\cite{Kimura:2011dc}. Combined with the assumptions that the quasi-static approximation holds and that the matter over-density is distributed spherically in space leads to the analytic form of \shc \ \citep[see Equation 3.15 in][]{Wright:2022krq}.  These assumptions in the derivation of \shc \ lead to a caveat: that in its current public state, {\tt Hi-COLA} is designed to work with theories that exhibit \textit{Vainshtein} screening. However, recent development of \HiCOLA has focused on extending the formalism to other screening mechanisms like K-mouflage, and these results are presented in \autoref{sec:kmouflage_results}. The full details of K-mouflage in {\tt Hi-COLA} will be the subject of an upcoming publication, Sen Gupta et al. \textit{(in prep.)}.


\subsubsection{COLA-FML} \label{sec:eftcola}

In this subsection we describe another approximate simulation method to modified gravity theories endowed with the Vainshtein mechanism, such as nDGP and the Galileon theory family. This method was initially proposed in Ref.~\cite{Scoccimarro:2009eu}, and later revisited in Ref.~\cite{Brando:2023fzu}. It consists of linearizing the Klein-Gordon equation in Fourier space, and implementing a resummation scheme to find a function, $G_{\rm eff}(k,a)$, defined in the same way as \autoref{eq:param_mg_evolution}, that approximately captures the nonlinear corrections introduced by the Vainshtein mechanism on small scales. Specifically, this function transitions between an unscreened regime at large scales, where $G_{\rm eff}(k,a) \to G_{\rm eff, L}(a)$, to the small scale regime where GR is recovered, $G_{\rm eff}(k,a) \to G_{\rm N}$.

Consequently, in order to do so, in Refs.~\cite{Scoccimarro:2009eu,Brando:2023fzu} the authors require the nonlinear function, $\Delta G_{\text {eff, NL}} (k,a)/G_{\rm N}$\footnote{We note that in Ref.~\cite{Brando:2023fzu} this function is called $M(k,a)$.}, has the screening property, i.e. 
\begin{align}\label{eq:M_asymp}
    \frac{\Delta G_{\text {eff, NL}} (k,a)}{G_{\rm N}}\left(k/k_{*} \ll 1 \right) & \to 1, \nonumber \\
    \frac{\Delta G_{\text {eff, NL}} (k,a)}{G_{\rm N}}\left(k/k_{*} \gg 1\right) & \to 0 \, ,
\end{align}
where $k_{*}$ is the wavenumber associated with the Vainshtein radius, defined in Eq.~\ref{eq:vain_r}.
The specifics behind the computation of the function $\Delta G_{\text {eff, NL}} (k,a)/G_{\rm N}$ is explicitly shown in Ref.~\cite{Brando:2023fzu}. This screening approximation scheme has the advantage of not introducing additional screening parameters used to tune the approximate results with results from \textit{N}-body simulations that consistently solve the full Klein-Gordon equation at each timestep of the simulation. The whole dependence of the gravity theory is encoded in the $\Delta G_{\text {eff, NL}} (k,a)/G_{\rm N}$ function.

The methodology of this approximate method for Vainshtein screening is computed using an external \texttt{Python} notebook, where one can follow the steps outlined in Ref.~\cite{Brando:2023fzu} to compute $G_{\rm eff}(k,a)$ externally. With the tabulated function computed, the results are then implemented in the \texttt{COLA-FML} (\href{https://github.com/HAWinther/FML/tree/master/FML/COLASolver}{\faicon{github}}) \textit{N}-body solver, that implements the COLA method in a parallelised manner, ideal for fast and approximate simulations. The \texttt{COLA-FML} library also has different screening approximations for theories other than the ones considered here, and are presented in Ref.~\cite{Winther:2014cia}. Importantly for this paper, our results for the $G_{\rm eff}(k,a)$ screening case will be different than the ones of \HiCOLA at nonlinear scales, however, at linear scales the two codes are identical.


\subsection{Halo Model Reaction} \label{sec:react}

The halo model reaction \citep{Cataneo:2018cic} is a flexible, accurate and fast means to model the nonlinear matter power spectrum. This model has been demonstrated to align with $N$-body simulations at the 2\% level down to $k=3 ~h / {\rm Mpc}$, with minor variations depending on redshift, the extent of modification to GR, and the mass of neutrinos \citep{Cataneo:2019fjp,Bose:2021mkz}. The method aims to model nonlinear corrections to the matter power spectrum resulting from modified gravity through the reaction $\mathcal{R}(k,z)$, which incorporates both 1-loop perturbation theory and the halo model \citep[see][for reviews]{Bernardeau:2001qr,Cooray:2002dia}. In this framework, the nonlinear matter power spectrum is expressed as the product
\begin{equation}
P_{\rm NL}(k,z) = \mathcal{R}(k,z)\, P^{\rm pseudo}_{\rm NL}(k,z) \, , 
\label{eq:nlpofkhmr}
\end{equation}
where the pseudo power spectrum is defined such that all nonlinear physics are modeled using GR but the initial conditions are adjusted to mimic the modified linear clustering at the target redshift. 

The halo model reaction without massive neutrinos, $\mathcal{R}(k,z)$, is given as a corrected ratio of target-to-pseudo halo model spectra 
\begin{equation}
\mathcal{R}(k,z) =  \frac{\{[1-\mathcal{E}(z)]\, e^{-k/k_\star(z)} + \mathcal{E}(z)\} \, P_{\rm 2H}(k,z)  +  P_{\rm 1H}(k,z)}{P_{\rm hm}^{\rm pseudo}(k,z)}\,. \label{eq:reaction}
\end{equation}
The components are explicitly given as 
\begin{align}
  P_{\rm hm}^{\rm pseudo}(k,z) = &   P_{\rm 2H} (k,z) + P_{\rm 1H}^{\rm pseudo}(k,z), \label{Pk-halos} \\ 
  \mathcal{E}(z) =& \lim_{k\rightarrow 0} \frac{ P_{\rm 1H}^{\rm }(k,z)}{ P_{\rm 1H}^{\rm pseudo}(k,z)} , \label{mathcale} \\ 
   k_{\rm \star}(z) = & - \bar{k} \left\{\ln \left[ 
    \frac{A(\bar{k},z)}{P_{\rm 2H}(\bar{k},z)} - \mathcal{E}(z) \right] - \ln\left[1-\mathcal{E}(z) \right]\right\}^{-1}\,, \label{kstar}
\end{align}
with
\begin{equation}
    A(k,z) =  \frac{P_{\rm 1-loop}(k,z)+ P_{\rm 1H}(k,z)}
    {P^{\rm pseudo}_{\rm 1-loop}(k,z)+ P_{\rm 1H}^{\rm pseudo}(k,z)}  P_{\rm hm}^{\rm pseudo}(k,z) -  P_{\rm 1H}(k,z)\,.
    \label{eq:sptcor}
\end{equation}
$P_{\rm 2H}(k,z)$ is the 2-halo term which we approximate with the linear matter power spectrum, $P_{\rm L}(k,z)$. $P_{\rm 1H}(k,z)$ and $P_{\rm 1H}^{\rm pseudo} (k,z)$ are the 1-halo terms as predicted by the halo model, with and without modifications to the standard spherical collapse equations, respectively.  Recall that by definition, the pseudo cosmology has no nonlinear beyond-$\Lambda$CDM modifications. Similarly, $P_{\rm 1-loop}(k,z)$ and $P_{\rm 1-loop}^{\rm pseudo} (k,z)$ are the standard perturbation theory 1-loop matter power spectra with and without nonlinear modifications to $\Lambda$CDM, respectively. As in the literature, \autoref{mathcale}'s limit is taken to be at $k=0.01\,h / {\rm Mpc}$ and $k_\star$ is computed using $\bar{k} = 0.06\,h /{\rm Mpc}$. 

The nDGP model was part of the initial release of the publicly available halo model reaction code, {\tt ReACT}~\citep{Bose:2020wch}. This code has been updated to include massive neutrinos in Ref.~\cite{Bose:2021mkz} and model independent parametrisations in Ref.~\cite{Bose:2022vwi}, which constituted version 2 of the code (\href{https://github.com/nebblu/ACTio-ReACTio}{\faicon{github}}). The GCCG model was recently implemented in this version of {\tt ReACT}~\citep{Atayde:2024tnr}, which is employed in \autoref{sec:results} in the CG limit. The K-mouflage patch is being made public with this work and we give all the relevant expressions in \autoref{sec:hmrkflage}.  

For the pseudo spectrum appearing in \autoref{eq:nlpofkhmr} we use {\tt HMCode2020}~\citep{Mead:2020vgs}. This is currently the most accurate and flexible prescription for the pseudo spectrum and has been tested in a number of works \citep[see for example][]{Cataneo:2018cic,Bose:2021mkz,Bose:2022vwi}. It is more accurate than the halofit prescription of Ref.~\cite{Takahashi:2012em}, quoting a $2.5-5\%$ accuracy for $k\leq 1~h/{\rm Mpc}$. It can also accommodate modifications that induce an additional scale dependence in the linear matter power spectrum. For modifications that only introduce a scale-independent shift in the linear spectrum amplitude, more accurate emulators can be used, such as the {\tt EuclidEmulator2}~\citep{Euclid:2018mlb}, which are quoted to be $1\%$ accurate when compared to high fidelity $N$-body simulations down to $k\leq 10~h/{\rm Mpc}$. Despite this, the reaction function $R(k,z)$ is only expected to be $1\%$ accurate for $k\leq 1~h/{\rm Mpc}$~\citep{Cataneo:2018cic}. 

It is also worth noting that {\tt EuclidEmulator2}'s internal accuracy is restricted to a hyperspheroidal region of their parameter space. Points outside this region might have considerable degradation in accuracy. This is considerably important in the context of beyond-$\Lambda$CDM scenarios as we need tools that work in extreme regions of the parameter space. For these reasons, work is currently being undertaken to build a pseudo spectrum emulator based on appropriate numerical simulations~\citep{Giblin:2019iit}. 

The choice of {\tt HMCode2020} keeps in line with the halo model reaction's claim of generality, while maintaining competitive accuracy within the reaction's percent-level accuracy range, especially when taking the ratio of modified to unmodified spectra, i.e. the matter power spectrum boost (see \autoref{eq:boost}). 


\section{Results} \label{sec:results}

Our main results are the comparisons of the nonlinear matter power spectrum between the different codes. Specifically, we consider the models described in \autoref{tab:sims} for which we have full $N$-body simulations available to use as benchmarks. We list the specifications of each simulation ran for these comparisons below. These include: box size ($L_{\rm box}$), number of particles ($N_{\rm P}$), particle mass ($m_{\rm P}$), grid cells ($N_{\rm g}$), initial redshift ($z_{\rm ini}$) and force resolution.
\begin{itemize}
    \item {\tt ECOSMOG} runs: 
    $L_{\rm box}=1024~\textrm{Mpc}/h$, \, $N_{\rm P}=1024^3$, \, $m_{\rm P}\simeq7.8\times10^{10}~M_\odot/h$. The initial conditions are generated at $z_{\rm ini}=49$ by {\tt MPGrafic}~\citep{Prunet:2008fv} using the Zel'dovich approximation. It uses a force resolution of  $\sim 15.6 \textrm{kpc}/h$.
    \item  
    {\tt MG-GLAM} runs: $L_{\rm box} = 512\,\mathrm{Mpc}/h, \, N_{\rm P} = 1024^3, \,  N_{\rm g} = 2048^3, \, m_{\rm P} = 1.07 \times 10^{10}\,M_{\odot},/h$, where $N_{\rm g}$ is the number of grid cells.  Initial conditions are generated at $ z_{\rm ini} = 100$  using {\tt GLAM}'s own initial condition generator. It uses a fixed force resolution of  $250\,\mathrm{kpc}/h$ with an adaptive time-stepping described in the original {\tt GLAM} paper~\citep{Klypin:2018MNRAS.478.4602K.GLAM}.
    \item 
    {\tt MG-evolution} runs: The nDGP simulation runs use  $L_{\rm box} = 1000 ~{\rm Mpc}/h$ with $N_{\rm g}=N_{\rm p} = 1024^3$. The initial conditions are generated at $z_{\rm ini} = 49$. 
    For the CG case, the initial conditions are the same as in the {\tt COLA} runs. These runs use $L_{\rm box}=400~{\rm Mpc}/h$, $N_{\rm P} = N_{\rm g}=512^3$.
    \item 
     {\tt COLA} runs: $L_{\rm box}=400~\textrm{Mpc}/h$ and $N_{\rm P}=512^3$ with initial conditions generated using 2LPT for all simulations. For K-mouflage: 
    $m_{\rm P}\simeq4.1\times10^{10}~{\rm M_\odot}/h$. Initial conditions are generated at $z_{\rm ini}=19$. For nDGP:
    $m_{\rm P}\simeq3.7\times10^{10}~M_\odot/h$. Initial conditions are generated at $z_{\rm ini}=49$. For CG and QCDM: 
    $m_{\rm P}\simeq4.1\times10^{10}~{\rm M_\odot}/h$. Initial conditions are generated at $z_{\rm ini}=49$.
\end{itemize}

Before presenting the spectra comparisons, we take a look at how each model presented in \autoref{sec:theory} modifies the standard $\Lambda$CDM background evolution. This background evolution is adopted for each of the different codes and so differences seen in the following section only arise from how the perturbations are treated. 

\begin{table*}
\centering
\caption{Models considered in this work. The $\Lambda$CDM $\sigma_8(z=0) = 0.851, 0.805, 0.815 $ for nDGP, CG and K-mouflage cosmologies respectively. We remind the reader that the values for $\{s,q\}$ are fixed in CG, while adopting the tracker solution for the scalar field imposes the values for $c_2$ and $c_3$ quoted in the table.}
\begin{tabular}{| c | c | c | c | c | c | c | c |   }
\hline  
 {\bf Parameter } & {\bf  nDGP-N1} & {\bf nDGP-N5} & {\bf CG} & {\bf QCDM} & {\bf K-mouflage - A} &  {\bf K-mouflage - B} &  {\bf K-mouflage - C}  \\ \hline 
$\Omega_{\rm m,0}$ &\multicolumn{2}{c}{0.281} &  \multicolumn{2}{c}{0.313} &  \multicolumn{3}{c}{0.3089}  \\
$\Omega_{\rm b,0}$ & \multicolumn{2}{c}{0.046}&  \multicolumn{2}{c}{0.049}  &  \multicolumn{3}{c}{0.0486} \\
$H_0$ & \multicolumn{2}{c}{69.7} &  \multicolumn{2}{c}{67.32}   &  \multicolumn{3}{c}{67.74} \\
$n_s$ &\multicolumn{2}{c}{0.971} &  \multicolumn{2}{c}{0.9655}  & \multicolumn{3}{c}{0.9667}  \\
$A_s$ & \multicolumn{2}{c}{$2.297\times10^{-9}$} &  \multicolumn{2}{c}{$2.010\times10^{-9}$}  & \multicolumn{3}{c}{$2.064\times10^{-9}$}  \\
 $\sigma_8(z=0)$ & 0.912 & 0.865 &  0.884  &  0.865  & 0.881 & 0.852 & 0.837  \\ \hline \hline 
$\Omega_{\rm rc}$ & 0.25  & 0.01  & -  &  -   & -   & - & - \\ \hline 
 $c_2/c_3^{2/3}$ & - & - & -5.378  &  -  & -   & - & -  \\
 $c_3$ & - & - & 10 &  -  & -   & - & -  \\  
 $s$ & - & - & 2.0 &  - & -   & -  & -  \\
 $q$ & - & - & 0.5 & - & -   & - & -  \\ \hline
  $n$ & - & - & - & -  & 2 & 2 &2  \\
 $\lambda$ & - & - & -  & - & 1.475 &  1.460 & 1.452  \\
 $K_0$ & - & - & -& -   & 1 & 10 & 1  \\
 $\beta_{\rm K}$ & - & -& -  & - & 0.2 & 0.2 & 0.1 \\
 \hline 
\end{tabular}
\label{tab:sims}
\end{table*}
\begin{figure*}
    \centering
    \includegraphics[width=\textwidth]{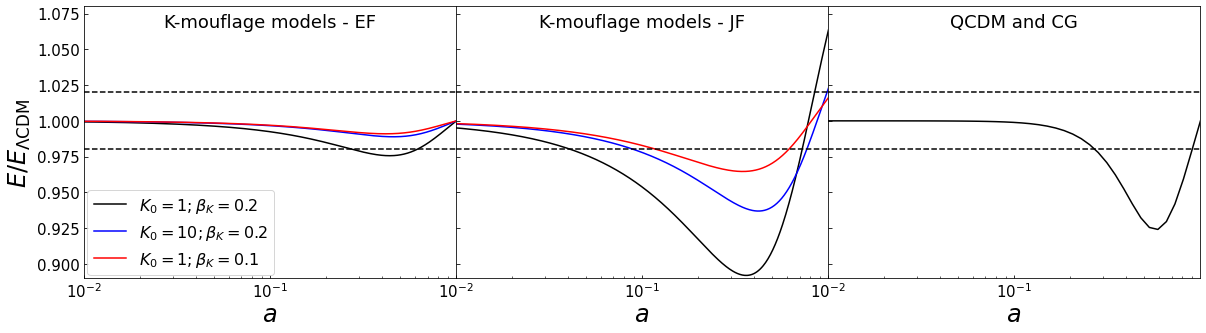}
    \vspace{-0.5cm}
    \caption{The ratio of the normalised Hubble expansion rate ($E(a) = H(a)/H_0$) between the modified gravity and GR models. The left panel shows the K-mouflage models shown in \autoref{tab:sims} in the Einstein frame, while the middle panel shows the same models in the Jordan frame, with $a$ now being the Jordan frame scale factor. The right panel shows the QCDM model, which has the same background expansion as the CG model. The model parameters for K-mouflage are defined in \autoref{sec:kmouflage}.}
    \label{fig:background}
\end{figure*}

\subsection{Background Evolution}

In \autoref{fig:background} we show the modification to the standard $\Lambda$CDM background expansion for the models described in \autoref{tab:sims}. We remind the reader that we assume a $\Lambda$CDM expansion for the nDGP models and so this is not shown. We see that the QCDM and CG cases give a much larger modification at late times than any of the K-mouflage models in the Einstein frame. In all models, we see a slower expansion rate at roughly $a>0.2$ which acts to enhance structure formation. Indeed, the $\sigma_8$ is larger for the  QCDM model than for both K-mouflage models B and C (see \autoref{tab:sims}), despite having a lower $A_s$ (although the QCDM cosmology has a slightly larger $\Omega_{\rm m,0}$). In all cases, the maximum modification is $\sim 8\%$ (QCDM), with the K-mouflage models giving a maximum modification of $3 \%$ at $a=0.5$. 

In the same figure we also show the modification in the Jordan frame for the K-mouflage models (middle panel). We see here that relative to $\Lambda$CDM, we have a significantly slower expansion at $a>0.03$, with a maxmimum modification of $11\%$ at $a \sim 0.4$. Further, the current day expansion is larger than the one expected from $\Lambda$CDM by $5\%$ under the strongest modification considered here. We do remind the reader that the free parameter $\lambda$ has been tuned to match the current day expansion rate in $\Lambda$CDM in the Einstein frame. These panels show that relatively large differences can be observed  at the background level when switching frames, which we will see in the next subsection are not evident at the level of the perturbations (also see \autoref{sec:frames}).

\subsection{Matter power spectrum boost}

Next we take a look at the perturbations, specifically how the matter power spectrum is modified over $\Lambda$CDM. For this, we consider the modified gravity boost, defined as 
\begin{equation}
    B(k,z) \equiv \frac{P_{\rm NL}(k,z)}{P^{\Lambda \rm CDM}_{\rm NL}(k,z)} \, . 
    \label{eq:boost}
\end{equation}
\begin{figure*}
    \centering
    \includegraphics[width=\textwidth]{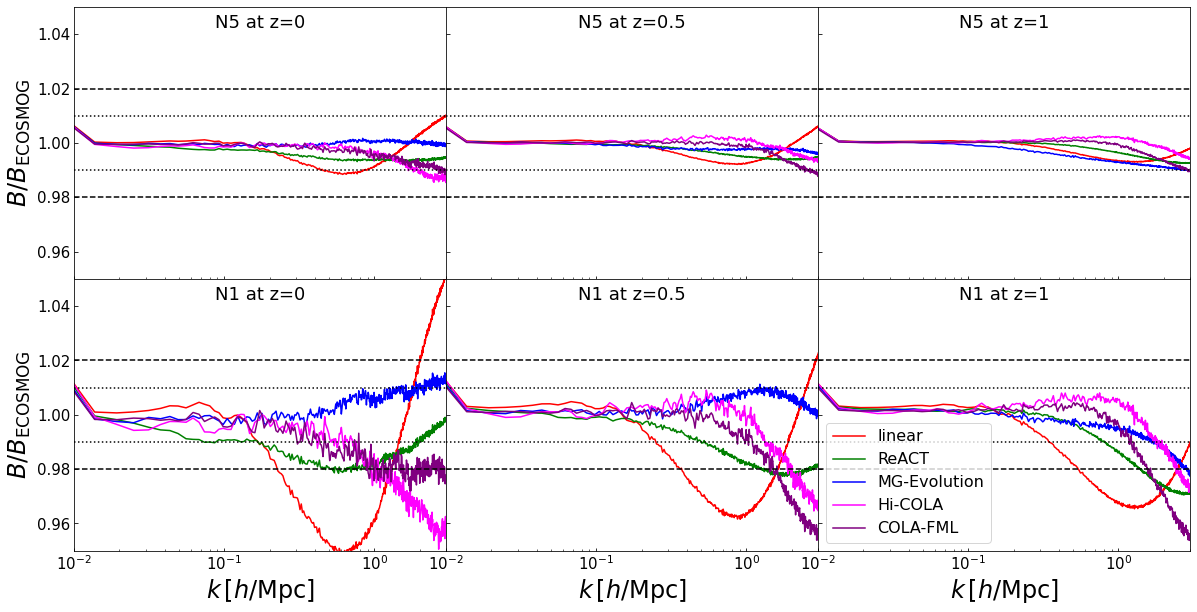}
    \caption{Comparing boost factors for the various codes listed in \autoref{tab:tools} for nDGP at $z=0,0.5,1$ (from left to right) with {\tt ECOSMOG} as benchmark. The upper panels show the results for the nDGP-N5 (low modification) model and the lower panels for the nDGP-N1 (high modification) model (see \autoref{tab:sims}). }
    \label{fig:dgp}
\end{figure*}
\subsubsection{nDGP}
In \autoref{fig:dgp} we show how the various predictions for the boost compare, using the {\tt ECOSMOG} measurements as a reference, for the nDGP cosmologies found in \autoref{tab:sims}. Boost comparisons for nDGP amongst different codes have already been performed extensively in the literature, and so this case is shown mainly as a consistency check, but also to compare the {\tt Hi-COLA} implementation which has not yet been tested before. 

We find that for the low modification case, N5, all predictions remain within $1\%$ of each other for $k\leq 3~h/{\rm Mpc}$, including the linear prediction, which for $z=0$ gives a modification of $B(k\rightarrow 0,z=0) = 1.033$. For N1, $B(k\rightarrow 0,z=0) = 1.149\%$. In this case, all predictions except the linear remain within $2\%$ of the {\tt ECOSMOG} reference for $k \leq 1~h/{\rm Mpc}$. {\tt MG-evolution} performs the best as expected, having an additional free parameter giving the screening transition, $k_*$. We have found $k_* = 2~h/{\rm Mpc}$ and $k_* =1~h/{\rm Mpc}$ give a good overall agreement with the {\tt ECOSMOG} simulations for the N1 and N5 models respectively.  Using these values the {\tt MG-evolution} boost remains within $1\%$ up to $k \leq 3~h/{\rm Mpc}$ except for the largest modes at $z=1$ where it worsens to $2\%$, consistent with what was found in Ref.~\cite{Hassani:2020rxd}. Similarly, the halo model reaction remains within $2\%$ for $k\leq 3~h/{\rm Mpc}$ except for the largest modes at $z=1$, where it worsens to $3\%$ agreement, in accordance with Ref.~\cite{Cataneo:2018cic}. 

The two COLA methods show similar agreement, but deviate the most on average from the reference boost measurements. {\tt COLA-FML} performs slightly better at $z=0$ while {\tt Hi-COLA} does better at higher $z$, with deviations up to $4\%$ at $k=3~h/{\rm Mpc}$. This is very consistent with the results of Ref.~\cite{Winther:2017jof}.  
%
%
\begin{figure*}
    \centering
    \includegraphics[width=\textwidth]{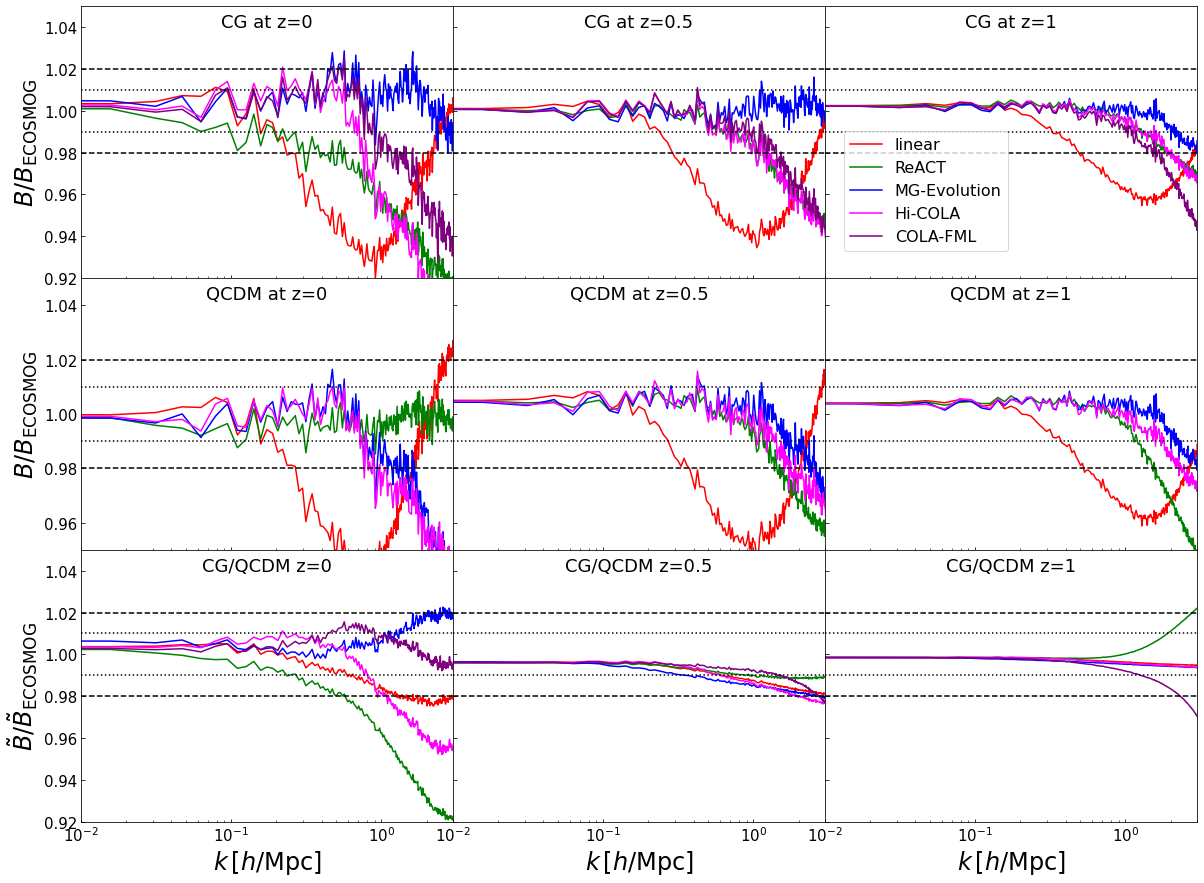}
    \caption{Comparing boost factors for the various codes listed in \autoref{tab:tools} for the CG (upper) and QCDM (middle) and QCDM-based boost (bottom) at $z=0,0.5,1$ (from left to right) with {\tt ECOSMOG} as the benchmark. Note {\tt COLA-FML} and {\tt Hi-COLA}'s results for QCDM are identical and so we only show the Hi-COLA QCDM ratio in the middle panels.}
    \label{fig:cg}
\end{figure*}

\begin{figure*}
    \centering
    \includegraphics[width=\textwidth]{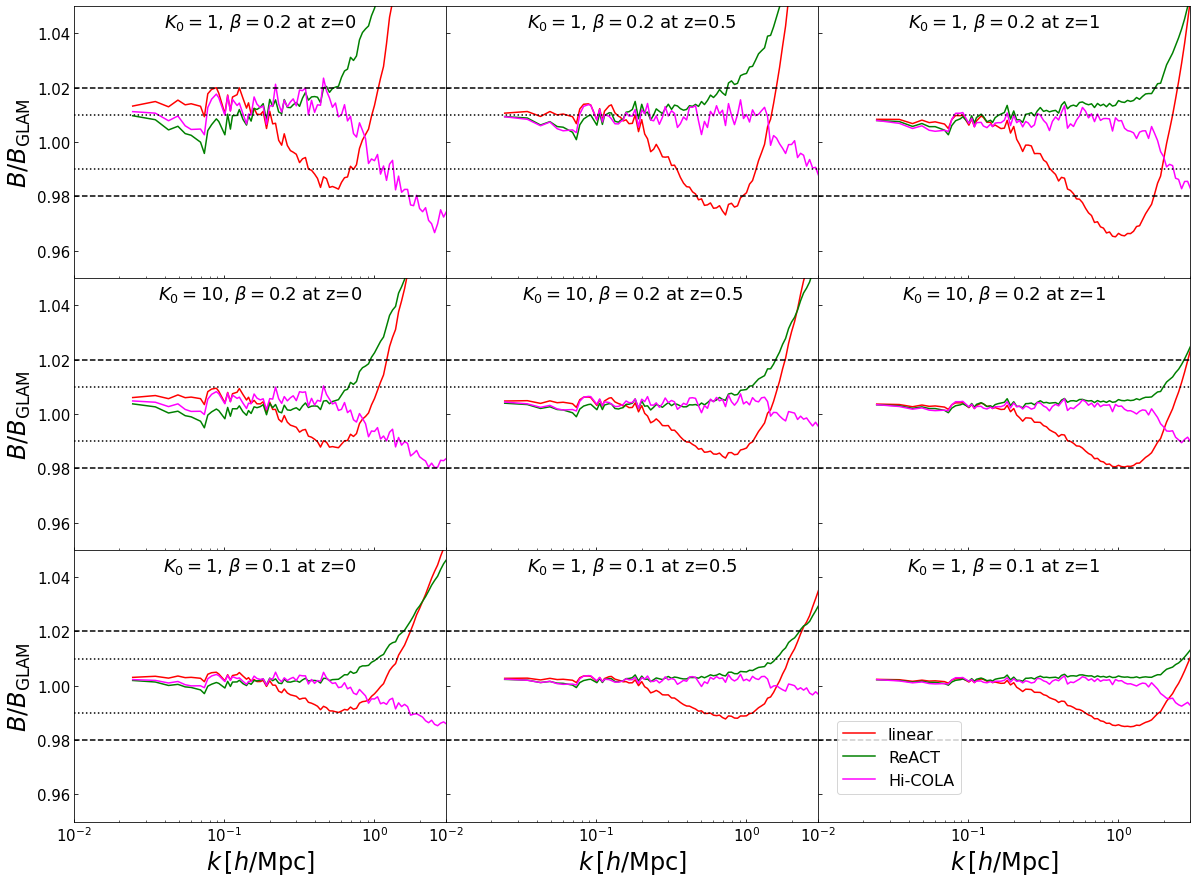}
    \caption{Comparing boost factors for the various codes listed in \autoref{tab:tools} for the K-mouflage models listed in \autoref{tab:sims} with $\{K_0, \beta_{\rm K} \} = \{1,0.2\}, \{10,0.2\}, \{1,0.1\}$  (from top to bottom) at $z=0,0.5,1$ (from left to right) with {\tt MG-GLAM} as the benchmark. All models assume $n=2$.}
    \label{fig:km}
\end{figure*}

\subsubsection{Cubic Galileon (CG)}
\autoref{fig:cg} shows the boost comparisons between the various codes for the CG and QCDM cases, again using {\tt ECOSMOG} as a reference. These {\tt ECOSMOG} simulations were ran using the same code as presented in Ref.~\cite{Barreira:2013eea}. We have changed the baseline cosmology for these new runs, particularly lowering the value of $A_s$ and $H_0$. We also run a $\Lambda$CDM counterpart with which to calculate \autoref{eq:boost}. Previous works have compared the boost ratio of the CG spectrum to that in QCDM \citep{Wright:2022krq}, or have performed direct spectra comparisons \citep{Atayde:2024tnr}. Further, in Ref.~\cite{Atayde:2024tnr} the authors found significant disagreement when using an {\tt HMCode2020} prescription, which was outperformed by the halofit pseudo spectrum prescription. This was being caused by the $\sigma_8$-dependent damping introduced into {\tt HMCode2020}~\citep{Mead:2020vgs}, which was not calibrated for particularly high values of $\sigma_8$ as that of the simulations found in Ref.~\cite{Barreira:2013eea}. The lower value of $\sigma_8$ in our simulations was found to greatly improve the performance of {\tt HMCode2020} over halofit. For comparison with previous work, we also show the comparisons for the ratio of CG to QCDM power spectra, or QCDM-based boost, in the bottom panels of \autoref{fig:cg}.

The {\tt MG-evolution} predictions again give the best agreement, remaining within $1\%$ in the CG case for $z\geq0.5$ down to $k=3~h/{\rm Mpc}$. 
For $z=1$, the linear implementation, or equivalently $k_* \to \infty$, provides the best match. However, in the figure, we have plotted the case $k_*(z\geq0.5) = 6~h/{\rm Mpc}$ as it appears to work well given the resolution of the simulation. Adopting the value $k_*=6~h/{\rm Mpc}$ at $z=0$ causes a quick divergence of the predictions as expected from the nature of $k_*$, amounting to a $8\%$ disagreement at $k=1~h/{\rm Mpc}$. Adopting $k_*=0.4~h/{\rm Mpc}$ at $z=0$ brings the predictions to within $2\%$ agreement in the same range of scales. Interestingly, in the QCDM-based boost case we can adopt the same value of $k_*=0.3~h/{\rm Mpc}$ for all redshifts considered while keeping a good fit to the {\tt ECOSMOG} measurements. This seems to indicate that $k_*$ is also degenerate to some extent with the background modification.

In the QCDM case, the predictions are consistent within $1\%$ down to $k=1~h/{\rm Mpc}$. The disagreement for  $k > 1~h/{\rm Mpc}$ arises from resolution effects, as evidenced by the agreement between { \tt MG-evolution} and { \tt Hi-COLA}. This suggests that the tuning of $k_*$ performed to match the reference boost factor in CG is compensating for the resolution-induced loss of boost due to the background by suppressing the small-scale screening.

The halo model reaction remains within $1\%$ for $k\leq 1~h/{\rm Mpc}$ for both QCDM and CG cases, with the exception of the CG at $z=0$. Here we find up to $4\%$ disagreement with {\tt ECOSMOG}. This is an atypically large disagreement given the similarity of CG to nDGP, for which the halo model reaction performs significantly better. To investigate this, we have tested different pseudo spectra prescriptions, specifically halofit and {\tt EuclidEmulator2}, neither offering significant improvement for the matter power spectrum boost. We have also tried omitting the 1-loop correction (see \autoref{eq:sptcor}) with little change to the predictions as found in Ref.~\cite{Bose:2022vwi}. The excellent agreement in the QCDM case at $z=0$, with $1\%$ agreement beyond $k=3~h/{\rm Mpc}$, indicates no issue in the background implementation. Further, the QCDM-based boost comparisons show the same disagreement at $z=0$, but the same or better agreement at higher redshifts, which is just a partial cancellation of inaccuracies in the QCDM and CG $\Lambda$CDM-based boost cases.

Lastly, we also checked the behaviour of the reaction function $\mathcal{R}$ for varying GCCG modification strengths by changing the value of $s$. We compared these to corresponding nDGP predictions for $\mathcal{R}$ such that the nDGP models gave the same linear enhancement of structure as the GCCG cases, making their pseudo spectra identical. We observed significantly more suppression coming from $\mathcal{R}$ in the GCCG than nDGP, especially for large modifications (large $s$ or large $\Omega_{\rm rc}$), which may be due to the $G_2$ term present in the GCCG. We do note that the CG has a very large linear enhancement of clustering at $z=0$, equivalent to a nDGP model with $\Omega_{\rm rc}=0.6$. This may indicate a break down of the halo model reaction's assumptions, specifically the $\Lambda$CDM fits it assumes for the halo mass function and virial concentration. The latter has been shown to significantly impact its accuracy \citep{Cataneo:2018cic,Srinivasan:2021gib,Srinivasan:2023qsu}, especially when the modification to gravity is large. To further pin the $z=0$ CG disagreement down, we would need to run a CG pseudo cosmology simulation which would make it clear whether or not the reaction modelling or $\Lambda$CDM-fits in the halo model components are failing. GCCG simulations with a smaller modification will also indicate this validity range. This will be the focus of future work. 

Finally, both COLA implementations remain $2\%$ consistent with {\tt ECOSMOG} in the CG case at scales $k\leq 1~h/{\rm Mpc}$. {\tt COLA-FML} performs slightly better at low $z$ while {\tt Hi-COLA} shows better agreement at higher $z$. The implementations differ only in their approach to screening and so we only show the {\tt Hi-COLA} results for QCDM, where it is  similarly consistent as in the CG case. We note that all codes tend to under-predict the boost at small scales. Part of this difference surely comes from the fact that while the {\tt ECOSMOG} code consistently solves the full Klein-Gordon equation, the other codes implement the screening mechanism in an approximate way, making use of the spherical approximation in one way or another. Therefore, at smaller scales these approximate methods are not guaranteed to be valid. A better test of the accuracy of screening is provided by the QCDM-based boost in the bottom panels, where we see far better agreement between the COLA methods and {\tt ECOSMOG}.

On this note, we remark that both COLA and {\tt MG-evolution}'s disagreement with the benchmarks in both nDGP, CG and QCDM cases is also partially due to a low force resolution which can lead to a loss of power on small scales~\citep[see][for example]{Brando:2022gvg}. By increasing the force resolution, and time steps in the COLA cases, we expect to find much better agreement above $k=1~h/{\rm Mpc}$, particularly in the QCDM case which does not have screening. We note that the limited force accuracy  will affect all particle mesh codes, including {\tt MG-GLAM}, and the most efficient and sure way to go to smaller scales would be to use Tree-particle mesh or AMR codes like {\tt ECOSMOG}. 

\begin{table*}
\centering
\caption{Maximal percent deviation of the nonlinear matter power spectrum boost under various modelling approaches against benchmarks, at different redshifts for $k\leq 1(3)~h/{\rm Mpc}$. }
\begin{tabular}{| c | c | c | c | c | c | c | c | c | c | c | c | c | }
\hline  
  & \multicolumn{3}{c}{{\tt MG-evolution}} & \multicolumn{3}{c}{{\tt COLA-FML}} & \multicolumn{3}{c}{{\tt Hi-COLA}} & \multicolumn{3}{c}{{\tt ReACT}}  \\ 
 Model & z=0 & z=0.5  & z=1 & z=0 & z=0.5 & z=1 & z=0 & z=0.5 & z=1 & z=0 & z=0.5 & z=1  \\ \hline
N1 & 1(1)\% &1(1)\% & 1(2)\% & 1(2)\% &1(4)\% & 1(4)\%& 2(4)\% &1(4)\% & 1(3)\% &  2(2)\% & 2(2) \%& 2(3)\% \\ 
N5 & 1(1)\% &1(1)\% & 1(1)\% & 1(1)\% &1(1)\% & 1(1)\% & 1(1)\% &1(1)\% & 1(1)\% & 1(1)\% & 1(1)\%&1(1)\%  \\ 
CG & 1(1)\% &1(1)\% & 1(2)\%& 1(6)\% & 1(5)\%& 1(5)\% & 1(10)\% &2(6)\% & 1(3)\% &  4(8)\% & 2(5)\% & 1(3)\% \\ 
QCDM & 1(5)\%& 1(3)\% & 1(2)\%&- &- & -&  2(6)\% & 1(3)\%& 1(3)\% & 1(1)\% & 1(4)\% & 1(5)\%  \\ 
K-mouflage A & - & - & - & - &- & -& 1(4)\% & 1(3)\%&1(3)\% & 2(17)\% & 1(8)\%&1(4)\% \\ 
K-mouflage B & -& - & -& - &- & -& 1(2)\% & 1(1)\%&1(1)\% & 2(10)\% & 1(5)\% & 1(2)\% \\ 
K-mouflage C & -&- & -&- &- & -& 1(1)\% & 1(1)\%& 1(1)\% & 1(4)\% & 1(3)\% & 1(1)\% \\ 
\end{tabular}
\label{tab:results}
\end{table*}

\subsubsection{K-mouflage}
\label{sec:kmouflage_results}
For K-mouflage, we restrict our comparisons to {\tt MG-GLAM}, {\tt Hi-COLA} and {\tt ReACT} with the {\tt MG-evolution} implementation to be the focus of an upcoming work. We expect the same level of accuracy as exhibited in the CG and nDGP cases, especially given the freedom imparted by $k_*$.  

In \autoref{fig:km} we show the results for the K-mouflage model. As a reference we use the {\tt MG-GLAM} simulations, ran for the purpose of this comparison. We compare the K-mouflage boost for the three models listed in \autoref{tab:sims}, all of which assume $n=2$ in \autoref{eq:kmnckinetic}. We begin by noting that the coupling of matter to the scalar field is proportional to $\beta_{\rm K}/K_0$~\citep[see Equation.~81 of][for example]{Brax:2014yla}, and so large positive $K_0$ decreases the fifth force while large $\beta_{\rm K}$ increases it. We find the larger the modification, the worse the agreement between {\tt ReACT}, {\tt Hi-COLA} and {\tt MG-GLAM}. We can see this by moving from top to bottom panels in \autoref{fig:km}. Further, we note for the largest modification (top panels), there is a $~1\%$ offset between {\tt MG-GLAM} and linear theory (as well as the other codes). This was also seen in Figure.~10 of Ref.~\cite{Hernandez-Aguayo:2021kuh} but not seen in the linearised simulations presented in that reference, suggesting this is a consequence of the nonlinear treatment of {\tt MG-GLAM}. We also note much smaller linear theory offsets at large scales for the weaker modifications.

For the strongest modification, K-mouflage A in \autoref{tab:sims}, at low $z$, all codes are consistent within $2\%$ for $k\leq1~h/{\rm Mpc}$. This agreement improves for the halo model reaction to $1\%$ agreement for $k\leq 3~h/{\rm Mpc}$ for $z=1$ and the weakest modification, K-mouflage C in \autoref{tab:sims}. Overall, {\tt Hi-COLA} does not show significant improvement or degradation with redshift or modification strength, consistently remaining within $2\%$ for $k\leq 3~h/{\rm Mpc}$. The exception is K-mouflage A for $z=0$ (upper left panel), where it degrades to 4\% discrepancy at $k=3~h/{\rm Mpc}$. The {\tt Hi-COLA} predictions are all made in the Jordan frame while {\tt MG-GLAM} and {\tt ReACT} produce predictions in the Einstein frame. It is here we note the consistency of the nonlinear matter power spectrum in both frames, confirming the claim of Ref.~\cite{Francfort:2019ynz}.

Before concluding we make some technical notes on the comparisons. In the case of the Jordan frame predictions from {\tt Hi-COLA}, the boost is taken with the K-mouflage spectrum measured at $a_{\rm J}$, calculated using \autoref{eq:scalefacframe}. Finally, we note that {\tt ReACT} has the option to use the PPF screening formalism for K-mouflage as derived in Ref.~\cite{Lombriser:2016zfz}, and which we present in \autoref{sec:ppfgeff} for completeness. This framework comes with an additional degree of freedom and so we have chosen not to use this in our comparisons. The inclusion of K-mouflage theories in {\tt Hi-COLA} will be the subject of an upcoming publication, Sen Gupta et al. \textit{(in prep.)}.
%

%

\section{Conclusions}
\label{sec:conclusions}
High quality $N$-body codes for modified gravity are essential in order to place reliable constraints on gravity using large-scale structure (LSS) observations. Ongoing galaxy surveys such as Euclid or the Dark Energy Survey will heighten their necessity by beating down the statistical uncertainty on our measurements, making theoretical accuracy essential. Benchmarking the accuracy of approximate but computationally efficient numerical methods against these high quality simulations is an important step towards reliable constraints from the forthcoming data. 

In this paper we have performed comparisons of the matter power spectrum modification induced by three distinct theories of modified gravity, each of which induces a scale-independent enhancement of the linear growth of structure: the normal branch of the DGP braneworld model, the Cubic Galileon  and K-mouflage. The former two employ the Vainshtein screening mechanism, while the latter employs the K-mouflage screening mechanism. For similar comparisons with scale-dependent modifications to the linear growth and the chameleon \citep{Khoury:2003aq} or 
symmmetron \citep{Hinterbichler:2010es} screening mechanisms, 
we refer the reader to Refs.~\cite{Winther:2015wla,Winther:2017jof,Cataneo:2018cic,Hassani:2020rxd}.

We compare the matter power spectrum boost predicted by 6 different numerical codes, each of which has a varying approach to the nonlinear gravitational coupling: full $N$-body ({\tt ECOSMOG} and {\tt MG-GLAM}), Comoving Lagrangian Acceleration (COLA) of which we compare two distinct codes, {\tt Hi-COLA} and {\tt COLA-FML}, the relativistic parametrised $N$-body code, {\tt MG-evolution}, and the semi-analytic halo model reaction approach expressed by the {\tt ReACT} code. We summarise the distinctions of each code below:
\begin{itemize}
    \item
    {\bf Full $N$-body}: Solves the Klein-Gordan equation exactly to get the force applied to particles in a box. Serves  as an accuracy benchmark.
    \item
    {\bf Hi-COLA}: Includes a fifth force in the COLA formalism via a screening factor, as well as consistently solving the modified cosmological expansion history. Screening factors are derived using a quasi-linear treatment of the metric and scalar field perturbations, along with assuming the quasi-static approximation and spherically distributed over-densities.
    \item
    {\bf COLA-FML}: Introduces the Vainshtein mechanism by evaluating a function, $G_{\rm eff}(k,a)$, that captures on average nonlinear corrections from the screening mechanism. This method is performed by linearizing the Klein-Gordon equation in Fourier space, and the full function is found by an iterative process.
    \item
    {\bf MG-evolution}: Employs a parametrised ansatz for the nonlinear force law which comes with a screening parameter. 
    \item 
    {\bf ReACT}: Uses spherical collapse, the halo model and 1-loop perturbation theory to predict the matter power spectrum. 
\end{itemize}
We summarize the overall accuracy exhibited by each approach in \autoref{tab:results} with respect to the full $N$-body benchmark. We remark that $N$-body codes solving the full Klein-Gordon equation in modified gravity are  $1\%$ consistent~\citep{Winther:2015wla} for $k\lesssim 7~h/{\rm Mpc}$ in their prediction for the boost.

We find that all approaches considered here are overall $2\%$ consistent with the benchmark $N$-body boost at scales $k\leq 1~h/{\rm Mpc}$ and for $z\leq 1$. The only exceptions are {\tt ReACT} for the strongest modifications to $\Lambda$CDM and at $z=0$. {\tt MG-evolution} performs the best, with a general accuracy of $1\%$ at all scales considered ($k\leq 3~h/{\rm Mpc}$), but this accuracy comes at the cost of tuning the screening parameter depending on the output redshift or modification strength, which might undermine the predictivity of the code. 

We thus can advocate the safe use of these codes, and any emulators based upon them~\citep[see][for example]{Tsedrik:2024cdi,Carrion:2024itc,Gordon:2024jaj}\footnote{The results of this work do not directly apply to the emulator produced in Ref.~\cite{Fiorini:2023fjl}, \texttt{nDGPemu}, as the screening approximation used to produced their training set is different from the ones adopted in this work.}, at fairly nonlinear scales for scale-independent models. We note the caveat that emulation error should be quantified and appropriately accounted for. 

For a more concrete estimate on the validity of these methods, we can consider a Euclid-like survey whose weak lensing analysis will have a signal to noise peaking at (conservatively) $z\approx0.7$~\citep[see][for example]{Euclid:2021rez}. Imposing a $2\%$ accuracy demand on the matter power spectrum model, and assuming a $\Lambda$CDM fiducial background cosmology, we can arguably trust all method predictions for $\ell_{\rm max}\lesssim 1800$. This roughly corresponds to the pessimistic scenario described in Ref.~\cite{Euclid:2019clj}.

At scales $k>1~h/{\rm Mpc}$ we find all codes begin to diverge by more than $2\%$ for the strongest modifications considered. They  should thus not be used to model the highly nonlinear scales of structure formation in the context of forthcoming LSS analyses without considering an appropriate theoretical error contribution to the error budget~\citep[see][for example]{Audren:2012vy}.    

The goal of this work was to validate different methods to compute the nonlinear matter power spectrum boost (see \autoref{eq:boost}). This function inherently depends on the nonlinear matter power spectrum of $\Lambda$CDM. But further, the boost must be applied to an accurate $\Lambda$CDM spectrum prediction in order to get a nonlinear modified matter power spectrum prediction. Therefore, the final modified gravity prescription inherits a dependence on predictions of the standard model. While we now have state-of-the-art high resolution tools to evaluate $P_{\rm NL}^{\rm \Lambda CDM}(k,z)$, the region in which these tools have internal accuracy within $1\%-2\%$ may not be as broad as we need for extracting unbiased constraints on cosmological parameters for Stage-IV LSS surveys~\citep[see][for a more in depth discussion]{Gordon:2024jaj}. Furhtermore it is expected that in beyond-$\Lambda$CDM analysis, extreme regions of the parameter space need to be sampled, which heightens the need for the development of more comprehensive emulators in $\Lambda$CDM, as these are also imperative for the study of beyond-$\Lambda$CDM models.

In a similar vein, a further investigation of the impact of baryons in a full parameter inference scenario remains imperative to perform using the codes validated in this work. It has been shown that the interplay between baryonic physics and cosmology exhibit some dependence at small scales~\citep{Elbers:2024dad}. However, it is unknown to what extent in the nonlinear regime we can still extract relevant cosmological information to improve our constraints, i.e., if we need to model baryonic physics deep inside the nonlinear regime, $k\sim 10 ~h/$Mpc or not. Alternatively to modelling baryonic physics at the level of the power spectrum, it would be interesting to investigate the performance of procedures that mitigate the impact of baryons in the parameter constrains, such as the methods described in Refs.~\cite{baryonic_pcas,baryon_pca,des_baryon_pca}.

To conclude, let us highlight that the methods compared in this work have been designed with an element of theoretical flexibility in mind. There is a general shift to move beyond hard-coded codes designed to run with only one gravity model, and instead build more general tools that can be calibrated to a range of different models\footnote{A caveat here is that a hard-coded full $N$-body code like {\tt MG-GLAM} or {\tt ECOSMOG} may always be needed for validation.}. This is an essential step forward to streamline the testing of new theoretical ideas with data from Stage IV surveys.


\section*{Acknowledgements}
\noindent B.B. is supported by a UKRI Stephen Hawking Fellowship (EP/W005654/2). 
A.S.G. is supported by a STFC PhD studentship. 
F.H. acknowledges the Research Council of Norway and the resources provided by 
UNINETT Sigma2 -- the National Infrastructure for High Performance Computing and 
Data Storage in Norway. F.H. is also supported by a grant from the Swiss National Supercomputing Centre (CSCS) under project ID s1051.
T.B. is supported by ERC Starting Grant SHADE (grant no.\,StG 949572) and by a Royal Society University Research Fellowship (grant no.\,URF$\backslash$R$\backslash$231006).
L.A. is supported by Fundação para a Ciência e a Tecnologia (FCT) through the research grants UIDB/04434/2020, UIDP/04434/2020 and  from the FCT PhD fellowship grant with ref. number 2022.11152.BD.
N.F. is supported by the Italian Ministry of University and Research (MUR) through the Rita Levi Montalcini project ``Tests of gravity on cosmic scales" with reference PGR19ILFGP.
L.A. and N.F.  also acknowledge the FCT project with ref. number PTDC/FIS-AST/0054/2021 and  the COST Action CosmoVerse, CA21136, supported by COST (European Cooperation in Science and Technology).
C.H-A. acknowledges support from the Excellence Cluster ORIGINS which is funded by the Deutsche Forschungsgemeinschaft (DFG, German Research Foundation) under Germany's Excellence Strategy -- EXC-2094 -- 390783311. 
G.B. is supported by the Alexander von Humboldt Foundation.
B.F. is supported by a Royal Society Enhancement Award (grant no. RF$\backslash$ERE$\backslash$210304).
This work used the DiRAC@Durham facility managed by the Institute for Computational Cosmology on behalf of the STFC DiRAC HPC Facility (www.dirac.ac.uk). The equipment was funded by BEIS capital funding via STFC capital grants ST/K00042X/1, ST/P002293/1, ST/R002371/1 and ST/S002502/1, Durham University and STFC operations grant ST/R000832/1. DiRAC is part of the National e-Infrastructure. 
\HiCOLA data was generated utilising Queen Mary's \href{https://doi.org/10.5281/zenodo.438045}{Apocrita HPC facility}, supported by QMUL Research-IT, and the Sciama High Performance Compute (HPC) cluster which is supported by the ICG, SEPNet and the University of Portsmouth. 
For the purpose of open access, the authors have applied a Creative Commons Attribution (CC BY) licence to any Author Accepted Manuscript version arising from this submission.

\section*{Data Availability}

The halo model reaction software used in this article is publicly available in the {\tt ACTio-ReACTio} repository at \url{https://github.com/nebblu/ACTio-ReACTio}. In the same repository we also provide a {\tt Mathematica} notebook, \href{https://github.com/nebblu/ACTio-ReACTio/tree/master/notebooks}{{\tt kmouflage.nb}}, which contains relevant calculations for the K-mouflage model. $N$-body matter power spectra measurements are available upon request.



\bibliographystyle{mnras}
\bibliography{main} %

\begin{thebibliography}{}
\makeatletter
\relax
\def\mn@urlcharsother{\let\do\@makeother \do\$\do\&\do\#\do\^\do\_\do\%\do\~}
\def\mn@doi{\begingroup\mn@urlcharsother \@ifnextchar [ {\mn@doi@}
  {\mn@doi@[]}}
\def\mn@doi@[#1]#2{\def\@tempa{#1}\ifx\@tempa\@empty \href
  {http://dx.doi.org/#2} {doi:#2}\else \href {http://dx.doi.org/#2} {#1}\fi
  \endgroup}
\def\mn@eprint#1#2{\mn@eprint@#1:#2::\@nil}
\def\mn@eprint@arXiv#1{\href {http://arxiv.org/abs/#1} {{\tt arXiv:#1}}}
\def\mn@eprint@dblp#1{\href {http://dblp.uni-trier.de/rec/bibtex/#1.xml}
  {dblp:#1}}
\def\mn@eprint@#1:#2:#3:#4\@nil{\def\@tempa {#1}\def\@tempb {#2}\def\@tempc
  {#3}\ifx \@tempc \@empty \let \@tempc \@tempb \let \@tempb \@tempa \fi \ifx
  \@tempb \@empty \def\@tempb {arXiv}\fi \@ifundefined
  {mn@eprint@\@tempb}{\@tempb:\@tempc}{\expandafter \expandafter \csname
  mn@eprint@\@tempb\endcsname \expandafter{\@tempc}}}

\bibitem[\protect\citeauthoryear{Abbott et~al.}{Abbott
  et~al.}{2016}]{DES:2016jjg}
Abbott T.,  et~al., 2016, \mn@doi [Mon. Not. Roy. Astron. Soc.]
  {10.1093/mnras/stw641}, 460, 1270

\bibitem[\protect\citeauthoryear{Abbott et~al.}{Abbott
  et~al.}{2017}]{Monitor:2017mdv}
Abbott B.~P.,  et~al., 2017, \mn@doi [Astrophys. J.]
  {10.3847/2041-8213/aa920c}, 848, L13

\bibitem[\protect\citeauthoryear{Adamek, Daverio, Durrer  \& Kunz}{Adamek
  et~al.}{2016a}]{Adamek:2016zes}
Adamek J.,  Daverio D.,  Durrer R.,   Kunz M.,  2016a, \mn@doi [JCAP]
  {10.1088/1475-7516/2016/07/053}, 07, 053

\bibitem[\protect\citeauthoryear{Adamek, Daverio, Durrer  \& Kunz}{Adamek
  et~al.}{2016b}]{Adamek:2015eda}
Adamek J.,  Daverio D.,  Durrer R.,   Kunz M.,  2016b, \mn@doi [Nature Phys.]
  {10.1038/nphys3673}, 12, 346

\bibitem[\protect\citeauthoryear{Aghanim et~al.}{Aghanim
  et~al.}{2020}]{Aghanim:2018eyx}
Aghanim N.,  et~al., 2020, \mn@doi [Astron. Astrophys.]
  {10.1051/0004-6361/201833910}, 641, A6

\bibitem[\protect\citeauthoryear{{Akeson} et~al.,}{{Akeson}
  et~al.}{2019}]{2019arXiv190205569A}
{Akeson} R.,  et~al., 2019, {The Wide Field Infrared Survey Telescope: 100
  Hubbles for the 2020s} (\mn@eprint {arXiv} {1902.05569})

\bibitem[\protect\citeauthoryear{Alam et~al.}{Alam
  et~al.}{2021}]{eBOSS:2020yzd}
Alam S.,  et~al., 2021, \mn@doi [Phys. Rev. D] {10.1103/PhysRevD.103.083533},
  103, 083533

\bibitem[\protect\citeauthoryear{Albrecht et~al.}{Albrecht
  et~al.}{2006}]{Albrecht:2006um}
Albrecht A.,  et~al., 2006, {Report of the Dark Energy Task Force} (\mn@eprint
  {arXiv} {astro-ph/0609591})

\bibitem[\protect\citeauthoryear{Albuquerque, Frusciante  \&
  Martinelli}{Albuquerque et~al.}{2022}]{Albuquerque:2021grl}
Albuquerque I.~S.,  Frusciante N.,   Martinelli M.,  2022, \mn@doi [Phys. Rev.
  D] {10.1103/PhysRevD.105.044056}, 105, 044056

\bibitem[\protect\citeauthoryear{Armendariz-Picon, Damour  \&
  Mukhanov}{Armendariz-Picon et~al.}{1999}]{Armendariz-Picon:1999hyi}
Armendariz-Picon C.,  Damour T.,   Mukhanov V.~F.,  1999, \mn@doi [Phys. Lett.
  B] {10.1016/S0370-2693(99)00603-6}, 458, 209

\bibitem[\protect\citeauthoryear{{Arnold}, {Leo}  \& {Li}}{{Arnold}
  et~al.}{2019}]{Arnold:2019NatAs...3..945A}
{Arnold} C.,  {Leo} M.,   {Li} B.,  2019, \mn@doi [Nature Astronomy]
  {10.1038/s41550-019-0823-y}, \href
  {https://ui.adsabs.harvard.edu/abs/2019NatAs...3..945A} {3, 945}

\bibitem[\protect\citeauthoryear{Arnold, Li, Giblin, Harnois-D\'eraps  \&
  Cai}{Arnold et~al.}{2021}]{Arnold:2021xtm}
Arnold C.,  Li B.,  Giblin B.,  Harnois-D\'eraps J.,   Cai Y.-C.,  2021, {}
  (\mn@eprint {arXiv} {2109.04984})

\bibitem[\protect\citeauthoryear{Atayde, Frusciante, Bose, Casas  \& Li}{Atayde
  et~al.}{2024}]{Atayde:2024tnr}
Atayde L.,  Frusciante N.,  Bose B.,  Casas S.,   Li B.,  2024, {Non-linear
  power spectrum and forecasts for Generalized Cubic Covariant Galileon}
  (\mn@eprint {arXiv} {2404.11471})

\bibitem[\protect\citeauthoryear{Audren, Lesgourgues, Bird, Haehnelt  \&
  Viel}{Audren et~al.}{2013}]{Audren:2012vy}
Audren B.,  Lesgourgues J.,  Bird S.,  Haehnelt M.~G.,   Viel M.,  2013,
  \mn@doi [JCAP] {10.1088/1475-7516/2013/01/026}, 01, 026

\bibitem[\protect\citeauthoryear{Babichev \& Deffayet}{Babichev \&
  Deffayet}{2013}]{Babichev:2013usa}
Babichev E.,  Deffayet C.,  2013, \mn@doi [Class. Quant. Grav.]
  {10.1088/0264-9381/30/18/184001}, 30, 184001

\bibitem[\protect\citeauthoryear{Babichev, Deffayet  \& Ziour}{Babichev
  et~al.}{2009}]{Babichev:2009ee}
Babichev E.,  Deffayet C.,   Ziour R.,  2009, \mn@doi [Int. J. Mod. Phys. D]
  {10.1142/S0218271809016107}, 18, 2147

\bibitem[\protect\citeauthoryear{Bag, Mishra  \& Sahni}{Bag
  et~al.}{2018}]{Bag:2018jle}
Bag S.,  Mishra S.~S.,   Sahni V.,  2018, \mn@doi [Phys. Rev. D]
  {10.1103/PhysRevD.97.123537}, 97, 123537

\bibitem[\protect\citeauthoryear{Baker, Bellini, Ferreira, Lagos, Noller  \&
  Sawicki}{Baker et~al.}{2017}]{Baker:2017hug}
Baker T.,  Bellini E.,  Ferreira P.~G.,  Lagos M.,  Noller J.,   Sawicki I.,
  2017, \mn@doi [Phys. Rev. Lett.] {10.1103/PhysRevLett.119.251301}, 119,
  251301

\bibitem[\protect\citeauthoryear{Baker, Clampitt, Jain  \& Trodden}{Baker
  et~al.}{2018}]{Baker:2018mnu}
Baker T.,  Clampitt J.,  Jain B.,   Trodden M.,  2018, \mn@doi [Phys. Rev. D]
  {10.1103/PhysRevD.98.023511}, 98, 023511

\bibitem[\protect\citeauthoryear{Baker et~al.}{Baker
  et~al.}{2022}]{LISACosmologyWorkingGroup:2022wjo}
Baker T.,  et~al., 2022, \mn@doi [JCAP] {10.1088/1475-7516/2022/08/031}, 08,
  031

\bibitem[\protect\citeauthoryear{Baker, Barausse, Chen, de Rham, Pieroni  \&
  Tasinato}{Baker et~al.}{2023}]{Baker:2022eiz}
Baker T.,  Barausse E.,  Chen A.,  de Rham C.,  Pieroni M.,   Tasinato G.,
  2023, \mn@doi [JCAP] {10.1088/1475-7516/2023/03/044}, 03, 044

\bibitem[\protect\citeauthoryear{Baldauf, Mirbabayi, Simonovi\'c  \&
  Zaldarriaga}{Baldauf et~al.}{2016}]{Baldauf:2016sjb}
Baldauf T.,  Mirbabayi M.,  Simonovi\'c M.,   Zaldarriaga M.,  2016, {LSS
  constraints with controlled theoretical uncertainties} (\mn@eprint {arXiv}
  {1602.00674})

\bibitem[\protect\citeauthoryear{Baldi \& Simpson}{Baldi \&
  Simpson}{2015}]{Baldi:2014ica}
Baldi M.,  Simpson F.,  2015, \mn@doi [Mon. Not. Roy. Astron. Soc.]
  {10.1093/mnras/stv405}, 449, 2239

\bibitem[\protect\citeauthoryear{Barreira, Li, Baugh  \& Pascoli}{Barreira
  et~al.}{2012}]{Barreira:2012kk}
Barreira A.,  Li B.,  Baugh C.~M.,   Pascoli S.,  2012, \mn@doi [Phys. Rev. D]
  {10.1103/PhysRevD.86.124016}, 86, 124016

\bibitem[\protect\citeauthoryear{Barreira, Li, Hellwing, Baugh  \&
  Pascoli}{Barreira et~al.}{2013a}]{Barreira:2013eea}
Barreira A.,  Li B.,  Hellwing W.~A.,  Baugh C.~M.,   Pascoli S.,  2013a,
  \mn@doi [JCAP] {10.1088/1475-7516/2013/10/027}, 10, 027

\bibitem[\protect\citeauthoryear{Barreira, Li, Baugh  \& Pascoli}{Barreira
  et~al.}{2013b}]{Barreira:2013xea}
Barreira A.,  Li B.,  Baugh C.~M.,   Pascoli S.,  2013b, \mn@doi [JCAP]
  {10.1088/1475-7516/2013/11/056}, 11, 056

\bibitem[\protect\citeauthoryear{Barreira, Li, Baugh  \& Pascoli}{Barreira
  et~al.}{2014}]{Barreira:2014jha}
Barreira A.,  Li B.,  Baugh C.,   Pascoli S.,  2014, \mn@doi [JCAP]
  {10.1088/1475-7516/2014/08/059}, 08, 059

\bibitem[\protect\citeauthoryear{Barreira, Brax, Clesse, Li  \&
  Valageas}{Barreira et~al.}{2015a}]{Barreira:2014gwa}
Barreira A.,  Brax P.,  Clesse S.,  Li B.,   Valageas P.,  2015a, \mn@doi
  [Phys. Rev. D] {10.1103/PhysRevD.91.063528}, 91, 063528

\bibitem[\protect\citeauthoryear{Barreira, Brax, Clesse, Li  \&
  Valageas}{Barreira et~al.}{2015b}]{Barreira:2015aea}
Barreira A.,  Brax P.,  Clesse S.,  Li B.,   Valageas P.,  2015b, \mn@doi
  [Phys. Rev. D] {10.1103/PhysRevD.91.123522}, 91, 123522

\bibitem[\protect\citeauthoryear{Barreira, S\'anchez  \& Schmidt}{Barreira
  et~al.}{2016}]{Barreira:2016ovx}
Barreira A.,  S\'anchez A.~G.,   Schmidt F.,  2016, \mn@doi [Phys. Rev. D]
  {10.1103/PhysRevD.94.084022}, 94, 084022

\bibitem[\protect\citeauthoryear{Barroso et~al.}{Barroso
  et~al.}{2024}]{Euclid:2024yrr}
Barroso J. A.~A.,  et~al., 2024, {Euclid. I. Overview of the Euclid mission}
  (\mn@eprint {arXiv} {2405.13491})

\bibitem[\protect\citeauthoryear{Battye, Pace  \& Trinh}{Battye
  et~al.}{2018}]{Battye:2018ssx}
Battye R.~A.,  Pace F.,   Trinh D.,  2018, \mn@doi [Phys. Rev. D]
  {10.1103/PhysRevD.98.023504}, 98, 023504

\bibitem[\protect\citeauthoryear{Becker, Arnold, Li  \& Heisenberg}{Becker
  et~al.}{2020}]{Becker:2020azq}
Becker C.,  Arnold C.,  Li B.,   Heisenberg L.,  2020, \mn@doi [JCAP]
  {10.1088/1475-7516/2020/10/055}, 10, 055

\bibitem[\protect\citeauthoryear{Belgacem, Finke, Frassino  \&
  Maggiore}{Belgacem et~al.}{2019}]{Belgacem:2018wtb}
Belgacem E.,  Finke A.,  Frassino A.,   Maggiore M.,  2019, \mn@doi [JCAP]
  {10.1088/1475-7516/2019/02/035}, 02, 035

\bibitem[\protect\citeauthoryear{Bellini et~al.}{Bellini
  et~al.}{2018}]{Bellini:2017avd}
Bellini E.,  et~al., 2018, \mn@doi [Phys. Rev. D] {10.1103/PhysRevD.97.023520},
  97, 023520

\bibitem[\protect\citeauthoryear{Benevento, Raveri, Lazanu, Bartolo, Liguori,
  Brax  \& Valageas}{Benevento et~al.}{2019}]{Benevento:2018xcu}
Benevento G.,  Raveri M.,  Lazanu A.,  Bartolo N.,  Liguori M.,  Brax P.,
  Valageas P.,  2019, \mn@doi [JCAP] {10.1088/1475-7516/2019/05/027}, 05, 027

\bibitem[\protect\citeauthoryear{Bernardeau, Colombi, Gaztanaga  \&
  Scoccimarro}{Bernardeau et~al.}{2002}]{Bernardeau:2001qr}
Bernardeau F.,  Colombi S.,  Gaztanaga E.,   Scoccimarro R.,  2002, \mn@doi
  [Phys. Rept.] {10.1016/S0370-1573(02)00135-7}, 367, 1

\bibitem[\protect\citeauthoryear{Bernardo, Bose, Franzmann, Hagstotz, He, Litsa
   \& Niedermann}{Bernardo et~al.}{2022}]{Bernardo:2022cck}
Bernardo H.,  Bose B.,  Franzmann G.,  Hagstotz S.,  He Y.,  Litsa A.,
  Niedermann F.,  2022, {} (\mn@eprint {arXiv} {2210.06810})

\bibitem[\protect\citeauthoryear{Blanchard et~al.}{Blanchard
  et~al.}{2020}]{Euclid:2019clj}
Blanchard A.,  et~al., 2020, \mn@doi [Astron. Astrophys.]
  {10.1051/0004-6361/202038071}, 642, A191

\bibitem[\protect\citeauthoryear{Bonici et~al.}{Bonici
  et~al.}{2023}]{Euclid:2022hdx}
Bonici M.,  et~al., 2023, \mn@doi [Astron. Astrophys.]
  {10.1051/0004-6361/202244445}, 670, A47

\bibitem[\protect\citeauthoryear{Bose \& Koyama}{Bose \&
  Koyama}{2016}]{Bose:2016qun}
Bose B.,  Koyama K.,  2016, \mn@doi [JCAP] {10.1088/1475-7516/2016/08/032}, 08,
  032

\bibitem[\protect\citeauthoryear{Bose, Baldi  \& Pourtsidou}{Bose
  et~al.}{2018}]{Bose:2017jjx}
Bose B.,  Baldi M.,   Pourtsidou A.,  2018, \mn@doi [JCAP]
  {10.1088/1475-7516/2018/04/032}, 04, 032

\bibitem[\protect\citeauthoryear{Bose, Cataneo, Tr\"oster, Xia, Heymans  \&
  Lombriser}{Bose et~al.}{2020}]{Bose:2020wch}
Bose B.,  Cataneo M.,  Tr\"oster T.,  Xia Q.,  Heymans C.,   Lombriser L.,
  2020, \mn@doi [Mon. Not. Roy. Astron. Soc.] {10.1093/mnras/staa2696}, 498,
  4650

\bibitem[\protect\citeauthoryear{Bose et~al.,}{Bose
  et~al.}{2021}]{Bose:2021mkz}
Bose B.,  et~al., 2021, \mn@doi [Mon. Not. Roy. Astron. Soc.]
  {10.1093/mnras/stab2731}, 508, 2479

\bibitem[\protect\citeauthoryear{Bose, Tsedrik, Kennedy, Lombriser, Pourtsidou
  \& Taylor}{Bose et~al.}{2022}]{Bose:2022vwi}
Bose B.,  Tsedrik M.,  Kennedy J.,  Lombriser L.,  Pourtsidou A.,   Taylor A.,
  2022, {} (\mn@eprint {arXiv} {2210.01094})

\bibitem[\protect\citeauthoryear{Brando, Falciano, Linder  \& Velten}{Brando
  et~al.}{2019}]{Brando:2019xbv}
Brando G.,  Falciano F.~T.,  Linder E.~V.,   Velten H. E.~S.,  2019, \mn@doi
  [JCAP] {10.1088/1475-7516/2019/11/018}, 11, 018

\bibitem[\protect\citeauthoryear{Brando, Fiorini, Koyama  \& Winther}{Brando
  et~al.}{2022}]{Brando:2022gvg}
Brando G.,  Fiorini B.,  Koyama K.,   Winther H.~A.,  2022, \mn@doi [JCAP]
  {10.1088/1475-7516/2022/09/051}, 09, 051

\bibitem[\protect\citeauthoryear{Brando, Koyama  \& Winther}{Brando
  et~al.}{2023}]{Brando:2023fzu}
Brando G.,  Koyama K.,   Winther H.~A.,  2023, \mn@doi [JCAP]
  {10.1088/1475-7516/2023/06/045}, 06, 045

\bibitem[\protect\citeauthoryear{Brax \& Valageas}{Brax \&
  Valageas}{2014a}]{Brax:2014wla}
Brax P.,  Valageas P.,  2014a, \mn@doi [Phys. Rev. D]
  {10.1103/PhysRevD.90.023507}, 90, 023507

\bibitem[\protect\citeauthoryear{Brax \& Valageas}{Brax \&
  Valageas}{2014b}]{Brax:2014yla}
Brax P.,  Valageas P.,  2014b, \mn@doi [Phys. Rev. D]
  {10.1103/PhysRevD.90.023508}, 90, 023508

\bibitem[\protect\citeauthoryear{Brax, van~de Bruck, Davis, Li  \& Shaw}{Brax
  et~al.}{2011}]{Brax:2011ja}
Brax P.,  van~de Bruck C.,  Davis A.-C.,  Li B.,   Shaw D.~J.,  2011, \mn@doi
  [Phys. Rev. D] {10.1103/PhysRevD.83.104026}, 83, 104026

\bibitem[\protect\citeauthoryear{Brax, Davis, Li, Winther  \& Zhao}{Brax
  et~al.}{2013}]{Brax:2013mua}
Brax P.,  Davis A.-C.,  Li B.,  Winther H.~A.,   Zhao G.-B.,  2013, \mn@doi
  [JCAP] {10.1088/1475-7516/2013/04/029}, 04, 029

\bibitem[\protect\citeauthoryear{Brax, Rizzo  \& Valageas}{Brax
  et~al.}{2015}]{Brax:2015lra}
Brax P.,  Rizzo L.~A.,   Valageas P.,  2015, \mn@doi [Phys. Rev. D]
  {10.1103/PhysRevD.92.043519}, 92, 043519

\bibitem[\protect\citeauthoryear{Brax, Casas, Desmond  \& Elder}{Brax
  et~al.}{2021}]{Brax:2021wcv}
Brax P.,  Casas S.,  Desmond H.,   Elder B.,  2021, \mn@doi [Universe]
  {10.3390/universe8010011}, 8, 11

\bibitem[\protect\citeauthoryear{Carrilho, Carrion, Bose, Pourtsidou, Hidalgo,
  Lombriser  \& Baldi}{Carrilho et~al.}{2022}]{Carrilho:2021rqo}
Carrilho P.,  Carrion K.,  Bose B.,  Pourtsidou A.,  Hidalgo J.~C.,  Lombriser
  L.,   Baldi M.,  2022, \mn@doi [Mon. Not. Roy. Astron. Soc.]
  {10.1093/mnras/stac641}, 512, 3691

\bibitem[\protect\citeauthoryear{Carrion, Carrilho, Spurio~Mancini, Pourtsidou
  \& Hidalgo}{Carrion et~al.}{2024}]{Carrion:2024itc}
Carrion K.,  Carrilho P.,  Spurio~Mancini A.,  Pourtsidou A.,   Hidalgo J.~C.,
  2024, {Dark Scattering: accelerated constraints from KiDS-1000 with
  $\tt{ReACT}$ and $\tt{CosmoPower}$} (\mn@eprint {arXiv} {2402.18562})

\bibitem[\protect\citeauthoryear{Casas et~al.}{Casas
  et~al.}{2023}]{Euclid:2023tqw}
Casas S.,  et~al., 2023, {Euclid: Constraints on f(R) cosmologies from the
  spectroscopic and photometric primary probes} (\mn@eprint {arXiv}
  {2306.11053})

\bibitem[\protect\citeauthoryear{Cataneo, Lombriser, Heymans, Mead, Barreira,
  Bose  \& Li}{Cataneo et~al.}{2019}]{Cataneo:2018cic}
Cataneo M.,  Lombriser L.,  Heymans C.,  Mead A.,  Barreira A.,  Bose S.,   Li
  B.,  2019, \mn@doi [Mon.\ Not.\ Roy.\ Astron.\ Soc.] {10.1093/mnras/stz1836},
  488, 2121

\bibitem[\protect\citeauthoryear{Cataneo, Emberson, Inman, Harnois-Deraps  \&
  Heymans}{Cataneo et~al.}{2020}]{Cataneo:2019fjp}
Cataneo M.,  Emberson J.~D.,  Inman D.,  Harnois-Deraps J.,   Heymans C.,
  2020, \mn@doi [Mon. Not. Roy. Astron. Soc.] {10.1093/mnras/stz3189}, 491,
  3101

\bibitem[\protect\citeauthoryear{Catena, Pietroni  \& Scarabello}{Catena
  et~al.}{2007}]{Catena:2006bd}
Catena R.,  Pietroni M.,   Scarabello L.,  2007, \mn@doi [Phys. Rev. D]
  {10.1103/PhysRevD.76.084039}, 76, 084039

\bibitem[\protect\citeauthoryear{Chiba \& Yamaguchi}{Chiba \&
  Yamaguchi}{2013}]{Chiba:2013mha}
Chiba T.,  Yamaguchi M.,  2013, \mn@doi [JCAP] {10.1088/1475-7516/2013/10/040},
  10, 040

\bibitem[\protect\citeauthoryear{Christiansen, Hassani, Jalilvand  \&
  Mota}{Christiansen et~al.}{2023}]{Christiansen:2023tfy}
Christiansen O.,  Hassani F.,  Jalilvand M.,   Mota D.~F.,  2023, \mn@doi
  [JCAP] {10.1088/1475-7516/2023/05/009}, 05, 009

\bibitem[\protect\citeauthoryear{Clifton, Carrilho, Fernandes  \&
  Mulryne}{Clifton et~al.}{2020}]{Clifton:2020xhc}
Clifton T.,  Carrilho P.,  Fernandes P. G.~S.,   Mulryne D.~J.,  2020, \mn@doi
  [Phys. Rev. D] {10.1103/PhysRevD.102.084005}, 102, 084005

\bibitem[\protect\citeauthoryear{Cooray \& Sheth}{Cooray \&
  Sheth}{2002}]{Cooray:2002dia}
Cooray A.,  Sheth R.~K.,  2002, \mn@doi [Phys. Rept.]
  {10.1016/S0370-1573(02)00276-4}, 372, 1

\bibitem[\protect\citeauthoryear{Creminelli \& Vernizzi}{Creminelli \&
  Vernizzi}{2017}]{Creminelli:2017sry}
Creminelli P.,  Vernizzi F.,  2017, \mn@doi [Phys. Rev. Lett.]
  {10.1103/PhysRevLett.119.251302}, 119, 251302

\bibitem[\protect\citeauthoryear{Creminelli, Lewandowski, Tambalo  \&
  Vernizzi}{Creminelli et~al.}{2018}]{Creminelli:2018xsv}
Creminelli P.,  Lewandowski M.,  Tambalo G.,   Vernizzi F.,  2018, \mn@doi
  [JCAP] {10.1088/1475-7516/2018/12/025}, 1812, 025

\bibitem[\protect\citeauthoryear{Davis, Li, Mota  \& Winther}{Davis
  et~al.}{2012}]{Davis:2011pj}
Davis A.-C.,  Li B.,  Mota D.~F.,   Winther H.~A.,  2012, \mn@doi [Astrophys.
  J.] {10.1088/0004-637X/748/1/61}, 748, 61

\bibitem[\protect\citeauthoryear{De~Felice \& Tsujikawa}{De~Felice \&
  Tsujikawa}{2012}]{DeFelice:2011bh}
De~Felice A.,  Tsujikawa S.,  2012, \mn@doi [JCAP]
  {10.1088/1475-7516/2012/02/007}, 02, 007

\bibitem[\protect\citeauthoryear{Deffayet, Esposito-Farese  \& Vikman}{Deffayet
  et~al.}{2009}]{Deffayet:2009wt}
Deffayet C.,  Esposito-Farese G.,   Vikman A.,  2009, \mn@doi [Phys. Rev. D]
  {10.1103/PhysRevD.79.084003}, 79, 084003

\bibitem[\protect\citeauthoryear{Deffayet, Gao, Steer  \& Zahariade}{Deffayet
  et~al.}{2011}]{Deffayet:2011gz}
Deffayet C.,  Gao X.,  Steer D.~A.,   Zahariade G.,  2011, \mn@doi [Phys. Rev.
  D] {10.1103/PhysRevD.84.064039}, 84, 064039

\bibitem[\protect\citeauthoryear{Dvali, Gabadadze  \& Porrati}{Dvali
  et~al.}{2000}]{Dvali:2000hr}
Dvali G.~R.,  Gabadadze G.,   Porrati M.,  2000, \mn@doi [Phys. Lett. B]
  {10.1016/S0370-2693(00)00669-9}, 485, 208

\bibitem[\protect\citeauthoryear{{Eifler}, {Krause}, {Dodelson}, {Zentner},
  {Hearin}  \& {Gnedin}}{{Eifler} et~al.}{2015}]{baryonic_pcas}
{Eifler} T.,  {Krause} E.,  {Dodelson} S.,  {Zentner} A.~R.,  {Hearin} A.~P.,
  {Gnedin} N.~Y.,  2015, \mn@doi [Mon. Notices Royal Astron. Soc.]
  {10.1093/mnras/stv2000}, \href
  {https://ui.adsabs.harvard.edu/abs/2015MNRAS.454.2451E} {454, 2451}

\bibitem[\protect\citeauthoryear{Elbers et~al.}{Elbers
  et~al.}{2024}]{Elbers:2024dad}
Elbers W.,  et~al., 2024, {The FLAMINGO project: the coupling between baryonic
  feedback and cosmology in light of the $S_8$ tension} (\mn@eprint {arXiv}
  {2403.12967})

\bibitem[\protect\citeauthoryear{Ezquiaga \& Zumalacárregui}{Ezquiaga \&
  Zumalacárregui}{2017}]{Ezquiaga:2017ekz}
Ezquiaga J.~M.,  Zumalacárregui M.,  2017, \mn@doi [Phys. Rev. Lett.]
  {10.1103/PhysRevLett.119.251304}, 119, 251304

\bibitem[\protect\citeauthoryear{Fiorini, Koyama  \& Baker}{Fiorini
  et~al.}{2023}]{Fiorini:2023fjl}
Fiorini B.,  Koyama K.,   Baker T.,  2023, \mn@doi [JCAP]
  {10.1088/1475-7516/2023/12/045}, 12, 045

\bibitem[\protect\citeauthoryear{Francfort, Ghosh  \& Durrer}{Francfort
  et~al.}{2019}]{Francfort:2019ynz}
Francfort J.,  Ghosh B.,   Durrer R.,  2019, \mn@doi [JCAP]
  {10.1088/1475-7516/2019/09/071}, 09, 071

\bibitem[\protect\citeauthoryear{Frusciante \& Pace}{Frusciante \&
  Pace}{2020}]{Frusciante:2020zfs}
Frusciante N.,  Pace F.,  2020, \mn@doi [Phys. Dark Univ.]
  {10.1016/j.dark.2020.100686}, 30, 100686

\bibitem[\protect\citeauthoryear{Frusciante \& Perenon}{Frusciante \&
  Perenon}{2020}]{Frusciante:2019xia}
Frusciante N.,  Perenon L.,  2020, \mn@doi [Phys. Rept.]
  {10.1016/j.physrep.2020.02.004}, 857, 1

\bibitem[\protect\citeauthoryear{Frusciante, Peirone, Atayde  \&
  De~Felice}{Frusciante et~al.}{2020}]{Frusciante:2019puu}
Frusciante N.,  Peirone S.,  Atayde L.,   De~Felice A.,  2020, \mn@doi [Phys.
  Rev. D] {10.1103/PhysRevD.101.064001}, 101, 064001

\bibitem[\protect\citeauthoryear{Frusciante et~al.}{Frusciante
  et~al.}{2023}]{Euclid:2023rjj}
Frusciante N.,  et~al., 2023, {Euclid: Constraining linearly scale-independent
  modifications of gravity with the spectroscopic and photometric primary
  probes} (\mn@eprint {arXiv} {2306.12368})

\bibitem[\protect\citeauthoryear{Gabadadze \& Iglesias}{Gabadadze \&
  Iglesias}{2006}]{Gabadadze:2006tf}
Gabadadze G.,  Iglesias A.,  2006, \mn@doi [Phys. Lett. B]
  {10.1016/j.physletb.2006.06.016}, 639, 88

\bibitem[\protect\citeauthoryear{Giblin, Cataneo, Moews  \& Heymans}{Giblin
  et~al.}{2019}]{Giblin:2019iit}
Giblin B.,  Cataneo M.,  Moews B.,   Heymans C.,  2019, \mn@doi [Mon. Not. Roy.
  Astron. Soc.] {10.1093/mnras/stz2659}, 490, 4826

\bibitem[\protect\citeauthoryear{Goldstein et~al.}{Goldstein
  et~al.}{2017}]{Goldstein:2017mmi}
Goldstein A.,  et~al., 2017, \mn@doi [Astrophys. J. Lett.]
  {10.3847/2041-8213/aa8f41}, 848, L14

\bibitem[\protect\citeauthoryear{Gordon, de Aguiar, Rebou\c{c}as, Brando,
  Falciano, Miranda, Koyama  \& Winther}{Gordon et~al.}{2024}]{Gordon:2024jaj}
Gordon J.,  de Aguiar B.~F.,  Rebou\c{c}as J.~a.,  Brando G.,  Falciano F.,
  Miranda V.,  Koyama K.,   Winther H.~A.,  2024, {Modeling nonlinear scales
  with COLA: preparing for LSST-Y1} (\mn@eprint {arXiv} {2404.12344})

\bibitem[\protect\citeauthoryear{Harnois-D\'eraps, Hernandez-Aguayo,
  Cuesta-Lazaro, Arnold, Li, Davies  \& Cai}{Harnois-D\'eraps
  et~al.}{2022}]{Harnois-Deraps:2022bie}
Harnois-D\'eraps J.,  Hernandez-Aguayo C.,  Cuesta-Lazaro C.,  Arnold C.,  Li
  B.,  Davies C.~T.,   Cai Y.-C.,  2022, {} (\mn@eprint {arXiv} {2211.05779})

\bibitem[\protect\citeauthoryear{Harry \& Noller}{Harry \&
  Noller}{2022}]{Harry:2022zey}
Harry I.,  Noller J.,  2022, \mn@doi [Gen. Rel. Grav.]
  {10.1007/s10714-022-03016-0}, 54, 133

\bibitem[\protect\citeauthoryear{Hassani \& Lombriser}{Hassani \&
  Lombriser}{2020}]{Hassani:2020rxd}
Hassani F.,  Lombriser L.,  2020, \mn@doi [Mon. Not. Roy. Astron. Soc.]
  {10.1093/mnras/staa2083}, 497, 1885

\bibitem[\protect\citeauthoryear{Hassani, Adamek, Kunz  \& Vernizzi}{Hassani
  et~al.}{2019}]{Hassani:2019lmy}
Hassani F.,  Adamek J.,  Kunz M.,   Vernizzi F.,  2019, \mn@doi [JCAP]
  {10.1088/1475-7516/2019/12/011}, 12, 011

\bibitem[\protect\citeauthoryear{Hassani, L'Huillier, Shafieloo, Kunz  \&
  Adamek}{Hassani et~al.}{2020}]{Hassani:2019wed}
Hassani F.,  L'Huillier B.,  Shafieloo A.,  Kunz M.,   Adamek J.,  2020,
  \mn@doi [JCAP] {10.1088/1475-7516/2020/04/039}, 04, 039

\bibitem[\protect\citeauthoryear{Hearin, Zentner  \& Ma}{Hearin
  et~al.}{2012}]{Hearin:2011bp}
Hearin A.~P.,  Zentner A.~R.,   Ma Z.,  2012, \mn@doi [JCAP]
  {10.1088/1475-7516/2012/04/034}, 04, 034

\bibitem[\protect\citeauthoryear{{Hern{\'a}ndez-Aguayo}, {Arnold}, {Li}  \&
  {Baugh}}{{Hern{\'a}ndez-Aguayo}
  et~al.}{2021}]{Hernandez-Aguayo:2021MNRAS.503.3867H}
{Hern{\'a}ndez-Aguayo} C.,  {Arnold} C.,  {Li} B.,   {Baugh} C.~M.,  2021,
  \mn@doi [\mnras] {10.1093/mnras/stab694}, \href
  {https://ui.adsabs.harvard.edu/abs/2021MNRAS.503.3867H} {503, 3867}

\bibitem[\protect\citeauthoryear{{Hern{\'a}ndez-Aguayo}, {Ruan}, {Li},
  {Arnold}, {Baugh}, {Klypin}  \& {Prada}}{{Hern{\'a}ndez-Aguayo}
  et~al.}{2022}]{Hernandez-Aguayo:2021kuh}
{Hern{\'a}ndez-Aguayo} C.,  {Ruan} C.-Z.,  {Li} B.,  {Arnold} C.,  {Baugh}
  C.~M.,  {Klypin} A.,   {Prada} F.,  2022, \mn@doi [\jcap]
  {10.1088/1475-7516/2022/01/048}, \href
  {https://ui.adsabs.harvard.edu/abs/2022JCAP...01..048H} {2022, 048}

\bibitem[\protect\citeauthoryear{Hildebrandt et~al.}{Hildebrandt
  et~al.}{2017}]{Hildebrandt:2016iqg}
Hildebrandt H.,  et~al., 2017, \mn@doi [Mon. Not. Roy. Astron. Soc.]
  {10.1093/mnras/stw2805}, 465, 1454

\bibitem[\protect\citeauthoryear{Hinterbichler \& Khoury}{Hinterbichler \&
  Khoury}{2010}]{Hinterbichler:2010es}
Hinterbichler K.,  Khoury J.,  2010, \mn@doi [Phys. Rev. Lett.]
  {10.1103/PhysRevLett.104.231301}, 104, 231301

\bibitem[\protect\citeauthoryear{Horndeski}{Horndeski}{1974}]{Horndeski:1974wa}
Horndeski G.~W.,  1974, \mn@doi [Int.J.Theor.Phys.] {10.1007/BF01807638}, 10,
  363

\bibitem[\protect\citeauthoryear{Hou, Bautista, Berti, Cuesta-Lazaro,
  Hern\'andez-Aguayo, Tr\"oster  \& Zheng}{Hou et~al.}{2023}]{Hou:2023kfp}
Hou J.,  Bautista J.,  Berti M.,  Cuesta-Lazaro C.,  Hern\'andez-Aguayo C.,
  Tr\"oster T.,   Zheng J.,  2023, \mn@doi [Universe]
  {10.3390/universe9070302}, 9, 302

\bibitem[\protect\citeauthoryear{Howlett, Manera  \& Percival}{Howlett
  et~al.}{2015}]{Howlett:2015hfa}
Howlett C.,  Manera M.,   Percival W.~J.,  2015, \mn@doi [Astron. Comput.]
  {10.1016/j.ascom.2015.07.003}, 12, 109

\bibitem[\protect\citeauthoryear{Hu \& Sawicki}{Hu \&
  Sawicki}{2007}]{Hu:2007nk}
Hu W.,  Sawicki I.,  2007, \mn@doi [Phys.Rev.] {10.1103/PhysRevD.76.064004},
  D76, 064004

\bibitem[\protect\citeauthoryear{Hu, Raveri, Frusciante  \& Silvestri}{Hu
  et~al.}{2014}]{Hu:2013twa}
Hu B.,  Raveri M.,  Frusciante N.,   Silvestri A.,  2014, \mn@doi [Phys. Rev.
  D] {10.1103/PhysRevD.89.103530}, 89, 103530

\bibitem[\protect\citeauthoryear{{Huang}, {Eifler}, {Mandelbaum}  \&
  {Dodelson}}{{Huang} et~al.}{2019}]{baryon_pca}
{Huang} H.-J.,  {Eifler} T.,  {Mandelbaum} R.,   {Dodelson} S.,  2019, \mn@doi
  [Mon. Not. of the Royal Astron. Soc.] {10.1093/mnras/stz1714}, \href
  {https://ui.adsabs.harvard.edu/abs/2019MNRAS.488.1652H} {488, 1652}

\bibitem[\protect\citeauthoryear{{Huang} et~al.,}{{Huang}
  et~al.}{2021}]{des_baryon_pca}
{Huang} H.-J.,  et~al., 2021, \mn@doi [Mon. Not. of the Royal Astron. Soc.]
  {10.1093/mnras/stab357}, \href
  {https://ui.adsabs.harvard.edu/abs/2021MNRAS.502.6010H} {502, 6010}

\bibitem[\protect\citeauthoryear{Ivezi\'c et~al.}{Ivezi\'c
  et~al.}{2019}]{LSST:2008ijt}
Ivezi\'c v.,  et~al., 2019, \mn@doi [Astrophys. J.] {10.3847/1538-4357/ab042c},
  873, 111

\bibitem[\protect\citeauthoryear{Jing}{Jing}{2005}]{Jing:2004fq}
Jing Y.~P.,  2005, \mn@doi [Astrophys. J.] {10.1086/427087}, 620, 559

\bibitem[\protect\citeauthoryear{Kable, Benevento, Frusciante, De~Felice  \&
  Tsujikawa}{Kable et~al.}{2022}]{Kable:2021yws}
Kable J.~A.,  Benevento G.,  Frusciante N.,  De~Felice A.,   Tsujikawa S.,
  2022, \mn@doi [JCAP] {10.1088/1475-7516/2022/09/002}, 09, 002

\bibitem[\protect\citeauthoryear{Khoury \& Weltman}{Khoury \&
  Weltman}{2004}]{Khoury:2003aq}
Khoury J.,  Weltman A.,  2004, \mn@doi [Phys. Rev. Lett.]
  {10.1103/PhysRevLett.93.171104}, 93, 171104

\bibitem[\protect\citeauthoryear{Khoury \& Wyman}{Khoury \&
  Wyman}{2009}]{Khoury:2009tk}
Khoury J.,  Wyman M.,  2009, \mn@doi [Phys. Rev. D]
  {10.1103/PhysRevD.80.064023}, 80, 064023

\bibitem[\protect\citeauthoryear{Kimura, Kobayashi  \& Yamamoto}{Kimura
  et~al.}{2012}]{Kimura:2011dc}
Kimura R.,  Kobayashi T.,   Yamamoto K.,  2012, \mn@doi [Phys. Rev. D]
  {10.1103/PhysRevD.85.024023}, 85, 024023

\bibitem[\protect\citeauthoryear{{Klypin} \& {Prada}}{{Klypin} \&
  {Prada}}{2018}]{Klypin:2018MNRAS.478.4602K.GLAM}
{Klypin} A.,  {Prada} F.,  2018, \mn@doi [\mnras] {10.1093/mnras/sty1340},
  \href {https://ui.adsabs.harvard.edu/abs/2018MNRAS.478.4602K} {478, 4602}

\bibitem[\protect\citeauthoryear{Knabenhans et~al.}{Knabenhans
  et~al.}{2019}]{Euclid:2018mlb}
Knabenhans M.,  et~al., 2019, \mn@doi [Mon. Not. Roy. Astron. Soc.]
  {10.1093/mnras/stz197}, 484, 5509

\bibitem[\protect\citeauthoryear{Kobayashi}{Kobayashi}{2019}]{Kobayashi:2019hrl}
Kobayashi T.,  2019, \mn@doi [Rept. Prog. Phys.] {10.1088/1361-6633/ab2429},
  82, 086901

\bibitem[\protect\citeauthoryear{Kobayashi, Yamaguchi  \& Yokoyama}{Kobayashi
  et~al.}{2010}]{Kobayashi:2010cm}
Kobayashi T.,  Yamaguchi M.,   Yokoyama J.,  2010, \mn@doi [Phys. Rev. Lett.]
  {10.1103/PhysRevLett.105.231302}, 105, 231302

\bibitem[\protect\citeauthoryear{Koyama \& Maartens}{Koyama \&
  Maartens}{2006}]{Koyama:2005kd}
Koyama K.,  Maartens R.,  2006, \mn@doi [JCAP] {10.1088/1475-7516/2006/01/016},
  01, 016

\bibitem[\protect\citeauthoryear{{LSST Dark Energy Science
  Collaboration}}{{LSST Dark Energy Science
  Collaboration}}{2012}]{Abate:2012za}
{LSST Dark Energy Science Collaboration} 2012, {Large Synoptic Survey
  Telescope: Dark Energy Science Collaboration} (\mn@eprint {arXiv}
  {1211.0310})

\bibitem[\protect\citeauthoryear{Lepori et~al.}{Lepori
  et~al.}{2022}]{Euclid:2021rez}
Lepori F.,  et~al., 2022, \mn@doi [Astron. Astrophys.]
  {10.1051/0004-6361/202142419}, 662, A93

\bibitem[\protect\citeauthoryear{{Levi} et~al.,}{{Levi}
  et~al.}{2019}]{Levi:2019ggs}
{Levi} M.,  et~al., 2019, in Bulletin of the American Astronomical Society.
  p.~57 (\mn@eprint {arXiv} {1907.10688})

\bibitem[\protect\citeauthoryear{Li, Zhao, Teyssier  \& Koyama}{Li
  et~al.}{2012}]{Li:2011vk}
Li B.,  Zhao G.-B.,  Teyssier R.,   Koyama K.,  2012, \mn@doi [JCAP]
  {10.1088/1475-7516/2012/01/051}, 01, 051

\bibitem[\protect\citeauthoryear{Li, Zhao  \& Koyama}{Li
  et~al.}{2013a}]{Li:2013nua}
Li B.,  Zhao G.-B.,   Koyama K.,  2013a, \mn@doi [JCAP]
  {10.1088/1475-7516/2013/05/023}, 05, 023

\bibitem[\protect\citeauthoryear{Li, Barreira, Baugh, Hellwing, Koyama, Pascoli
   \& Zhao}{Li et~al.}{2013b}]{Li:2013tda}
Li B.,  Barreira A.,  Baugh C.~M.,  Hellwing W.~A.,  Koyama K.,  Pascoli S.,
  Zhao G.-B.,  2013b, \mn@doi [JCAP] {10.1088/1475-7516/2013/11/012}, 11, 012

\bibitem[\protect\citeauthoryear{Llinares, Mota  \& Winther}{Llinares
  et~al.}{2014}]{Llinares:2013jza}
Llinares C.,  Mota D.~F.,   Winther H.~A.,  2014, \mn@doi [Astron. Astrophys.]
  {10.1051/0004-6361/201322412}, 562, A78

\bibitem[\protect\citeauthoryear{Lombriser}{Lombriser}{2016}]{Lombriser:2016zfz}
Lombriser L.,  2016, \mn@doi [JCAP] {10.1088/1475-7516/2016/11/039}, 11, 039

\bibitem[\protect\citeauthoryear{Lombriser \& Lima}{Lombriser \&
  Lima}{2017}]{Lombriser:2016yzn}
Lombriser L.,  Lima N.~A.,  2017, \mn@doi [Phys. Lett.]
  {10.1016/j.physletb.2016.12.048}, B765, 382

\bibitem[\protect\citeauthoryear{Lombriser \& Taylor}{Lombriser \&
  Taylor}{2016}]{Lombriser:2015sxa}
Lombriser L.,  Taylor A.,  2016, \mn@doi [JCAP]
  {10.1088/1475-7516/2016/03/031}, 1603, 031

\bibitem[\protect\citeauthoryear{Lombriser, Hu, Fang  \& Seljak}{Lombriser
  et~al.}{2009}]{Lombriser:2009xg}
Lombriser L.,  Hu W.,  Fang W.,   Seljak U.,  2009, \mn@doi [Phys. Rev. D]
  {10.1103/PhysRevD.80.063536}, 80, 063536

\bibitem[\protect\citeauthoryear{Lue}{Lue}{2006}]{Lue:2005ya}
Lue A.,  2006, \mn@doi [Phys. Rept.] {10.1016/j.physrep.2005.10.007}, 423, 1

\bibitem[\protect\citeauthoryear{Luty, Porrati  \& Rattazzi}{Luty
  et~al.}{2003}]{Luty:2003vm}
Luty M.~A.,  Porrati M.,   Rattazzi R.,  2003, \mn@doi [JHEP]
  {10.1088/1126-6708/2003/09/029}, 09, 029

\bibitem[\protect\citeauthoryear{Martinelli et~al.}{Martinelli
  et~al.}{2021}]{Euclid:2020tff}
Martinelli M.,  et~al., 2021, \mn@doi [Astron. Astrophys.]
  {10.1051/0004-6361/202039835}, 649, A100

\bibitem[\protect\citeauthoryear{Mead, Heymans, Lombriser, Peacock, Steele  \&
  Winther}{Mead et~al.}{2016}]{Mead:2016zqy}
Mead A.,  Heymans C.,  Lombriser L.,  Peacock J.,  Steele O.,   Winther H.,
  2016, \mn@doi [Mon. Not. Roy. Astron. Soc.] {10.1093/mnras/stw681}, 459, 1468

\bibitem[\protect\citeauthoryear{Mead, Brieden, Tr\"oster  \& Heymans}{Mead
  et~al.}{2020}]{Mead:2020vgs}
Mead A.,  Brieden S.,  Tr\"oster T.,   Heymans C.,  2020, \mn@doi [Mon. Not.
  Roy. Astron. Soc.] {10.1093/mnras/stab082}

\bibitem[\protect\citeauthoryear{Nicolis, Rattazzi  \& Trincherini}{Nicolis
  et~al.}{2009}]{Nicolis:2008in}
Nicolis A.,  Rattazzi R.,   Trincherini E.,  2009, \mn@doi [Phys. Rev. D]
  {10.1103/PhysRevD.79.064036}, 79, 064036

\bibitem[\protect\citeauthoryear{Nouri-Zonoz, Hassani  \& Kunz}{Nouri-Zonoz
  et~al.}{2024}]{Nouri-Zonoz:2024dph}
Nouri-Zonoz A.~R.,  Hassani F.,   Kunz M.,  2024, {$k$-e$\mu$lator: emulating
  clustering effects of the $k$-essence dark energy} (\mn@eprint {arXiv}
  {2405.10424})

\bibitem[\protect\citeauthoryear{Pace, Battye, Bellini, Lombriser, Vernizzi  \&
  Bolliet}{Pace et~al.}{2021}]{Pace:2020qpj}
Pace F.,  Battye R.,  Bellini E.,  Lombriser L.,  Vernizzi F.,   Bolliet B.,
  2021, \mn@doi [JCAP] {10.1088/1475-7516/2021/06/017}, 06, 017

\bibitem[\protect\citeauthoryear{Peirone, Frusciante, Hu, Raveri  \&
  Silvestri}{Peirone et~al.}{2018}]{Peirone:2017vcq}
Peirone S.,  Frusciante N.,  Hu B.,  Raveri M.,   Silvestri A.,  2018, \mn@doi
  [Phys. Rev. D] {10.1103/PhysRevD.97.063518}, 97, 063518

\bibitem[\protect\citeauthoryear{Peirone, Benevento, Frusciante  \&
  Tsujikawa}{Peirone et~al.}{2019}]{Peirone:2019aua}
Peirone S.,  Benevento G.,  Frusciante N.,   Tsujikawa S.,  2019, \mn@doi
  [Phys. Rev. D] {10.1103/PhysRevD.100.063540}, 100, 063540

\bibitem[\protect\citeauthoryear{Perlmutter et~al.}{Perlmutter
  et~al.}{1999}]{SupernovaCosmologyProject:1998vns}
Perlmutter S.,  et~al., 1999, \mn@doi [Astrophys. J.] {10.1086/307221}, 517,
  565

\bibitem[\protect\citeauthoryear{Piga, Marinucci, D'Amico, Pietroni, Vernizzi
  \& Wright}{Piga et~al.}{2023}]{Piga:2022mge}
Piga L.,  Marinucci M.,  D'Amico G.,  Pietroni M.,  Vernizzi F.,   Wright
  B.~S.,  2023, \mn@doi [JCAP] {10.1088/1475-7516/2023/04/038}, 04, 038

\bibitem[\protect\citeauthoryear{Prunet, Pichon, Aubert, Pogosyan, Teyssier  \&
  Gottloeber}{Prunet et~al.}{2008}]{Prunet:2008fv}
Prunet S.,  Pichon C.,  Aubert D.,  Pogosyan D.,  Teyssier R.,   Gottloeber S.,
   2008, \mn@doi [Astrophys. J. Suppl.] {10.1086/590370}, 178, 179

\bibitem[\protect\citeauthoryear{Puchwein, Baldi  \& Springel}{Puchwein
  et~al.}{2013}]{Puchwein:2013lza}
Puchwein E.,  Baldi M.,   Springel V.,  2013, \mn@doi [Mon. Not. Roy. Astron.
  Soc.] {10.1093/mnras/stt1575}, 436, 348

\bibitem[\protect\citeauthoryear{Quartin, Tsujikawa, Amendola  \&
  Sturani}{Quartin et~al.}{2023}]{Quartin:2023tpl}
Quartin M.,  Tsujikawa S.,  Amendola L.,   Sturani R.,  2023, \mn@doi [JCAP]
  {10.1088/1475-7516/2023/08/049}, 08, 049

\bibitem[\protect\citeauthoryear{Ramachandra, Valogiannis, Ishak  \&
  Heitmann}{Ramachandra et~al.}{2021}]{Ramachandra:2020lue}
Ramachandra N.,  Valogiannis G.,  Ishak M.,   Heitmann K.,  2021, \mn@doi
  [Phys. Rev. D] {10.1103/PhysRevD.103.123525}, 103, 123525

\bibitem[\protect\citeauthoryear{Renk, Zumalac\'arregui, Montanari  \&
  Barreira}{Renk et~al.}{2017}]{Renk:2017rzu}
Renk J.,  Zumalac\'arregui M.,  Montanari F.,   Barreira A.,  2017, \mn@doi
  [JCAP] {10.1088/1475-7516/2017/10/020}, 10, 020

\bibitem[\protect\citeauthoryear{Riess et~al.}{Riess
  et~al.}{1998}]{SupernovaSearchTeam:1998fmf}
Riess A.~G.,  et~al., 1998, \mn@doi [Astron. J.] {10.1086/300499}, 116, 1009

\bibitem[\protect\citeauthoryear{{Ruan}, {Hern{\'a}ndez-Aguayo}, {Li},
  {Arnold}, {Baugh}, {Klypin}  \& {Prada}}{{Ruan} et~al.}{2022}]{Ruan:2021wup}
{Ruan} C.-Z.,  {Hern{\'a}ndez-Aguayo} C.,  {Li} B.,  {Arnold} C.,  {Baugh}
  C.~M.,  {Klypin} A.,   {Prada} F.,  2022, \mn@doi [\jcap]
  {10.1088/1475-7516/2022/05/018}, \href
  {https://ui.adsabs.harvard.edu/abs/2022JCAP...05..018R} {2022, 018}

\bibitem[\protect\citeauthoryear{Sakstein \& Jain}{Sakstein \&
  Jain}{2017}]{Sakstein:2017xjx}
Sakstein J.,  Jain B.,  2017, \mn@doi [Phys. Rev. Lett.]
  {10.1103/PhysRevLett.119.251303}, 119, 251303

\bibitem[\protect\citeauthoryear{Sawicki \& Bellini}{Sawicki \&
  Bellini}{2015}]{Sawicki:2015zya}
Sawicki I.,  Bellini E.,  2015, \mn@doi [Phys. Rev. D]
  {10.1103/PhysRevD.92.084061}, 92, 084061

\bibitem[\protect\citeauthoryear{Schmidt}{Schmidt}{2009a}]{Schmidt:2009sg}
Schmidt F.,  2009a, \mn@doi [Phys. Rev. D] {10.1103/PhysRevD.80.043001}, 80,
  043001

\bibitem[\protect\citeauthoryear{Schmidt}{Schmidt}{2009b}]{Schmidt:2009sv}
Schmidt F.,  2009b, \mn@doi [Phys. Rev. D] {10.1103/PhysRevD.80.123003}, 80,
  123003

\bibitem[\protect\citeauthoryear{Schmidt, Hu  \& Lima}{Schmidt
  et~al.}{2010}]{Schmidt:2009yj}
Schmidt F.,  Hu W.,   Lima M.,  2010, \mn@doi [Phys. Rev. D]
  {10.1103/PhysRevD.81.063005}, 81, 063005

\bibitem[\protect\citeauthoryear{Scoccimarro}{Scoccimarro}{2009}]{Scoccimarro:2009eu}
Scoccimarro R.,  2009, \mn@doi [Phys. Rev. D] {10.1103/PhysRevD.80.104006}, 80,
  104006

\bibitem[\protect\citeauthoryear{Sefusatti, Crocce, Scoccimarro  \&
  Couchman}{Sefusatti et~al.}{2016}]{Sefusatti:2015aex}
Sefusatti E.,  Crocce M.,  Scoccimarro R.,   Couchman H.,  2016, \mn@doi [Mon.
  Not. Roy. Astron. Soc.] {10.1093/mnras/stw1229}, 460, 3624

\bibitem[\protect\citeauthoryear{Simpson}{Simpson}{2010}]{Simpson:2010vh}
Simpson F.,  2010, \mn@doi [Phys. Rev. D] {10.1103/PhysRevD.82.083505}, 82,
  083505

\bibitem[\protect\citeauthoryear{Srinivasan, Thomas, Pace  \&
  Battye}{Srinivasan et~al.}{2021}]{Srinivasan:2021gib}
Srinivasan S.,  Thomas D.~B.,  Pace F.,   Battye R.,  2021, \mn@doi [JCAP]
  {10.1088/1475-7516/2021/06/016}, 06, 016

\bibitem[\protect\citeauthoryear{Srinivasan, Thomas  \& Battye}{Srinivasan
  et~al.}{2024}]{Srinivasan:2023qsu}
Srinivasan S.,  Thomas D.~B.,   Battye R.,  2024, \mn@doi [JCAP]
  {10.1088/1475-7516/2024/03/039}, 03, 039

\bibitem[\protect\citeauthoryear{Takahashi, Sato, Nishimichi, Taruya  \&
  Oguri}{Takahashi et~al.}{2012}]{Takahashi:2012em}
Takahashi R.,  Sato M.,  Nishimichi T.,  Taruya A.,   Oguri M.,  2012, \mn@doi
  [Astrophys. J.] {10.1088/0004-637X/761/2/152}, 761, 152

\bibitem[\protect\citeauthoryear{Tassev, Zaldarriaga  \& Eisenstein}{Tassev
  et~al.}{2013}]{Tassev:2013pn}
Tassev S.,  Zaldarriaga M.,   Eisenstein D.,  2013, \mn@doi [JCAP]
  {10.1088/1475-7516/2013/06/036}, 06, 036

\bibitem[\protect\citeauthoryear{Teyssier}{Teyssier}{2002}]{Teyssier:2001cp}
Teyssier R.,  2002, \mn@doi [Astron. Astrophys.] {10.1051/0004-6361:20011817},
  385, 337

\bibitem[\protect\citeauthoryear{Tsedrik, Bose, Carrilho, Pourtsidou, Pamuk,
  Casas  \& Lesgourgues}{Tsedrik et~al.}{2024}]{Tsedrik:2024cdi}
Tsedrik M.,  Bose B.,  Carrilho P.,  Pourtsidou A.,  Pamuk S.,  Casas S.,
  Lesgourgues J.,  2024, {Stage-IV Cosmic Shear with Modified Gravity and
  Model-independent Screening} (\mn@eprint {arXiv} {2404.11508})

\bibitem[\protect\citeauthoryear{Vainshtein}{Vainshtein}{1972}]{Vainshtein:1972sx}
Vainshtein A.,  1972, \mn@doi [Phys.Lett.] {10.1016/0370-2693(72)90147-5}, B39,
  393

\bibitem[\protect\citeauthoryear{Valogiannis \& Bean}{Valogiannis \&
  Bean}{2017}]{Valogiannis:2016ane}
Valogiannis G.,  Bean R.,  2017, \mn@doi [Phys. Rev. D]
  {10.1103/PhysRevD.95.103515}, 95, 103515

\bibitem[\protect\citeauthoryear{Will}{Will}{2014}]{Will:2014kxa}
Will C.~M.,  2014, \mn@doi [Living Rev. Rel.] {10.12942/lrr-2014-4}, 17, 4

\bibitem[\protect\citeauthoryear{Winther \& Ferreira}{Winther \&
  Ferreira}{2015a}]{Winther:2014cia}
Winther H.~A.,  Ferreira P.~G.,  2015a, \mn@doi [Phys. Rev. D]
  {10.1103/PhysRevD.91.123507}, 91, 123507

\bibitem[\protect\citeauthoryear{Winther \& Ferreira}{Winther \&
  Ferreira}{2015b}]{Winther:2015pta}
Winther H.~A.,  Ferreira P.~G.,  2015b, \mn@doi [Phys. Rev. D]
  {10.1103/PhysRevD.92.064005}, 92, 064005

\bibitem[\protect\citeauthoryear{Winther et~al.}{Winther
  et~al.}{2015}]{Winther:2015wla}
Winther H.~A.,  et~al., 2015, \mn@doi [Mon. Not. Roy. Astron. Soc.]
  {10.1093/mnras/stv2253}, 454, 4208

\bibitem[\protect\citeauthoryear{Winther, Koyama, Manera, Wright  \&
  Zhao}{Winther et~al.}{2017}]{Winther:2017jof}
Winther H.~A.,  Koyama K.,  Manera M.,  Wright B.~S.,   Zhao G.-B.,  2017,
  \mn@doi [JCAP] {10.1088/1475-7516/2017/08/006}, 08, 006

\bibitem[\protect\citeauthoryear{Wright, Sen~Gupta, Baker, Valogiannis  \&
  Fiorini}{Wright et~al.}{2023}]{Wright:2022krq}
Wright B.~S.,  Sen~Gupta A.,  Baker T.,  Valogiannis G.,   Fiorini B.,  2023,
  \mn@doi [JCAP] {10.1088/1475-7516/2023/03/040}, 03, 040

\bibitem[\protect\citeauthoryear{Zhao}{Zhao}{2014}]{Zhao:2013dza}
Zhao G.-B.,  2014, \mn@doi [Astrophys. J. Suppl.] {10.1088/0067-0049/211/2/23},
  211, 23

\bibitem[\protect\citeauthoryear{Zumalac\'arregui, Bellini, Sawicki,
  Lesgourgues  \& Ferreira}{Zumalac\'arregui
  et~al.}{2017}]{Zumalacarregui:2016pph}
Zumalac\'arregui M.,  Bellini E.,  Sawicki I.,  Lesgourgues J.,   Ferreira
  P.~G.,  2017, \mn@doi [JCAP] {10.1088/1475-7516/2017/08/019}, 08, 019

\bibitem[\protect\citeauthoryear{de Rham \& Melville}{de~Rham \&
  Melville}{2018}]{deRham:2018red}
de Rham C.,  Melville S.,  2018, \mn@doi [Phys. Rev. Lett.]
  {10.1103/PhysRevLett.121.221101}, 121, 221101

\bibitem[\protect\citeauthoryear{de Rham, Melville  \& Noller}{de~Rham
  et~al.}{2021}]{deRham:2021fpu}
de Rham C.,  Melville S.,   Noller J.,  2021, \mn@doi [JCAP]
  {10.1088/1475-7516/2021/08/018}, 08, 018

\makeatother
\end{thebibliography}



\appendix


\section{K-mouflage ReACT patch} \label{sec:hmrkflage}

Here we present the expressions needed to calculate the halo model reaction (see \autoref{eq:reaction}) in the K-mouflage model. The halo model reaction  relies on both the halo model~\citep[see][for a review]{Cooray:2002dia} and 1-loop perturbation theory~\citep[see][for a review]{Bernardeau:2001qr}. In particular, besides the background expansion $H(a)$, we require the modifications to the 1st, 2nd, 3rd order perturbative and nonlinear Poisson equations, as well as contributions to the potential energy of halos in order to solve the virial theorem~\citep[see][for more details]{Cataneo:2018cic,Bose:2020wch}. K-mouflage also comes with a friction term correction to the Euler equation~\citep{Brax:2014yla}.  

\subsection{Background}
For the background expansion we must solve the Klein-Gordon and Friedmann equations simultaneously. We do this numerically in {\tt ReACT} as done in Ref.~\cite{Hernandez-Aguayo:2021kuh}. The Friedmann equations are 
\begin{align}
    \frac{\HH^2}{H_0^2}\left[1-\frac{\varphi'{}^2}{6} \right]  = & \frac{A(\varphi) \Omega_{\rm m,0}}{a} + \frac{1}{3} \lambda^2 a^2  \nonumber \\ 
    & + \frac{(2n - 1)}{3} \lambda^2 a^2 K_0 \left(\frac{\varphi'{}^2}{2 \lambda^2 a^2} \right)^n \frac{\HH^{2n}}{H_0^{2n}} \, , \label{eq:fried1km} \\ 
    \frac{{\rm d}\HH}{{\rm d \tau}} \frac{1}{H_0^2} = & -\frac{A(\varphi) \Omega_{\rm m,0}}{2a} + \frac{1}{3} \lambda^2 a^2 - \frac{1}{3} \varphi'{}^2 \frac{\HH^2}{H_0^2} \nonumber \\ 
    & - \frac{(n + 1)}{3} \lambda^2 a^2 K_0 \left(\frac{\varphi'{}^2}{2 \lambda^2 a^2} \right)^n \frac{\HH^{2n}}{H_0^{2n}}  \, , \label{eq:fried2km}
\end{align}
while the Klein-Gordan equation is given as
\begin{align}
    & (K_X + 2 \bar{X} K_{XX}) \left[\frac{\HH^2}{H_0^2} \varphi'' + \frac{{\rm d} \HH }{{\rm d \tau}} \frac{1}{H_0^2} \varphi' \right] \nonumber \\ 
    & +2(K_X - \bar{X}K_{XX})\frac{\HH^2}{H_0^2}\varphi' + 3 \frac{{\rm d} A(\varphi)}{\rm d \varphi} \frac{\Omega_{\rm m,0}}{a} = 0  \, , \label{eq:kgkm}
\end{align}
 where $K_{XX} = {\rm d}^2 K/ {\rm d} X^2$ and we recall primes denote derivatives with respect to $\ln a$. In these equations we have defined the normalised scalar field $\varphi \equiv \phi/M_{\rm pl}$ and used the conformal Hubble rate $\HH(a) = H(a) a$. $\tau$ is conformal time. We indicate that a few typographical errors did occur in Ref.~\cite{Hernandez-Aguayo:2021kuh} which have been corrected in the above equations. 

To solve these equations we first find the analytic solutions to \autoref{eq:fried1km} for a given value of $n$~\footnote{We provide a \href{https://github.com/nebblu/ACTio-ReACTio/tree/master/notebooks}{\texttt{Mathematica} notebook} which computes the solutions for $n=2,3$ and checks consistency of the equations.}. For $n=2$ this is a quadratic equation in $\HH^2/H_0^2$. Then, for a given value of $a$ (or $\ln a$) we can substitute $\HH$ and \autoref{eq:fried2km} in \autoref{eq:kgkm}, enabling us to solve for the entire evolution of $\varphi$ (and $\varphi'$), and consequently $H(a)$.

\subsection{Perturbations}
The linear modification to the Poisson equation is given by \citep{Brax:2014yla}
\begin{equation}\label{eq:geff_lin_kmou}
\frac{G_{\rm eff,L}}{G_{\rm N}} = A(\varphi) \left( 1 +  \frac{2 \beta_{\rm K}^2}{K_X} \right) \, . 
\end{equation}
Here we have included the conformal factor $A(\varphi)$, that comes along with $\rho_{\rm m}$ in the Poisson equation, \autoref{eq:poi_mg}. Note that $G_{\rm eff,L}/G_{\rm N} = \mu(k,a)$ in the {\tt ReACT} standard notation of Refs.~\cite{Bose:2016qun,Cataneo:2018cic,Bose:2020wch,Bose:2022vwi} for example.

The 2nd and 3rd order symmetrised modifications to the Poisson equation, in the same notation of Ref.~\cite{Bose:2016qun,Bose:2020wch}, are \citep{Brax:2014yla}
\begin{align}
    \gamma_2(\bfk_1, \bfk_2, a) = & 0 \,, \nonumber \\ 
    \gamma_3(\bfk_1,\bfk_2, \bfk_3, a) = & -\frac{9}{2} K_{XX} \left(\frac{A(\varphi) \Omega_{\rm m,0}}{a} \frac{H_0^2}{\HH^2}\right)^3 \left(\frac{\beta_{\rm K}}{K_X}\right)^4 \frac{\HH^4}{H_0^2} \frac{1}{a^2 \lambda^2} \nonumber \\ 
    & \times \left[ \frac{(\mu_{12} + 2 \mu_{13} \mu_{23})}{k_1 k_2} + \frac{(\mu_{13} + 2 \mu_{23} \mu_{12})}{k_1 k_3} \nonumber \right. \\
    & \left. + \frac{(\mu_{23} + 2 \mu_{13} \mu_{12})}{k_2 k_3} \right] \, ,
\end{align}
where we have defined $\mu_{ij} \equiv \hat{k}_{i} \cdot \hat{k}_j$ and $k_i = |\bfk_i|$. 

Lastly, we also have a modification to the Euler equation in the form of a friction term~\citep{Brax:2014yla}. Similar terms have been included in {\tt ReACT} in the context of interacting dark energy models~\citep{Simpson:2010vh,Baldi:2014ica,Bose:2017jjx,Carrilho:2021rqo}. In the K-mouflage model considered here, this term is given as 
\begin{equation}
A_{\rm friction} = \beta_{\rm K} \varphi' \, . \label{eq:kmfric}
\end{equation}
This term enters the Euler equation as expressed in Equation.~2.10 of Ref.~\cite{Bose:2017jjx} for example.

\subsection{Spherical collapse}

The halo model reaction also requires us to solve for the spherical top-hat overdensity. This involves solving the evolution equation for the top-hat radius which requires specification of the nonlinear Poisson equation. The modification to this equation is to a good approximation equal to the linear modification at extra galactic scales given the smallness of the K-mouflage radius~\citep{Brax:2014yla}
\begin{equation}
\frac{G_{\text {eff}} (k,a)}{G_{\rm N}} = \frac{G_{\text {eff, L}} (a)}{G_{\rm N}} \, .  \label{eq:nlforcekm}
\end{equation}
We note in the notation of Ref.~\cite{Cataneo:2018cic},  $\mathcal{F} = G_{\rm eff}/G_{\rm N} - 1 = \Delta G_{\rm eff}/G_{\rm N}$. In {\tt ReACT} $\mathcal{F}$ appears as $1+\mathcal{F}$ in the Poisson equation. This yields the correct conformal factor accounting for the Einstein-frame transformation of the background density in the nonlinear Poisson equation, as it is already explicit in \autoref{eq:geff_lin_kmou}.

Lastly, we note that the top-hat radius evolution also must include the friction term \autoref{eq:kmfric}. 

\subsection{Virial theorem}
Here we present the potential energy contributions to the virial theorem in the K-mouflage model considered. This is needed to calculate the virial mass in the halo model reaction calculations. The specific components we require are~\citep{Schmidt:2009yj,Cataneo:2018cic}
\begin{align}
    \frac{W_{\rm N}}{E_0} & = - \Omega_{\rm m,0} \frac{a^{-1}}{a_{\rm i}^2} y^2 (1+\delta) \, ;   \\ 
    \frac{W_{\rm \phi}}{E_0} & = - \Omega_{\rm m,0} \mathcal{F} \,  \frac{a^{-1}}{a_{\rm i}^2} y^2  \delta \, ; \label{eq:phicontvir}  \\ 
    \frac{W_{\rm eff}}{E_0} & = - \frac{1}{3 M_{\rm pl}^2 H_0^2} (1+3w_{\rm eff}) \, \bar{\rho}_{\rm eff}  \frac{a^2}{a_{\rm i}^2} y^2  \, ;   \label{eq:weffvt} \\ 
    \frac{W_{\rm fric}}{E_0} & = -2 A_{\rm friction} \frac{H^2}{H_0^2}\frac{a^2}{a_{\rm i}^2} \, y \,  y' \, , 
\end{align}
where $y \equiv \frac{R_{\rm TH}}{R_{\rm i}}\frac{a_{\rm i}}{a}$, $R_{\rm TH}$ being the comoving top-hat radius, $R_{\rm i}$ the initial top-hat radius and $E_0$ is a normalisation. These represent the Newtonian contribution, the scalar field contribution, the effective dark energy contribution and a frictional force contribution as derived in Ref.~\cite{Carrilho:2021rqo}. In the K-mouflage model the scalar field affects both force enhancement and acts as an effective dark energy component. 

We recall that $\mathcal{F} = G_{\rm eff}/G_{\rm N} - 1$ which does not account for the correct conformal factor to appear in \autoref{eq:phicontvir} in the K-mouflage model. In this case we should have 
\begin{align}
\frac{W_{\rm \phi}}{E_0}  &= - \Omega_{\rm m,0} \left[ G_{\rm eff,L}/G_{\rm N} - A(\varphi) \right] \,  \frac{a^{-1}}{a_{\rm i}^2} y^2  \delta \, , \nonumber \\ 
& = - \Omega_{\rm m,0} \left[ A(\varphi) \frac{2 \beta_{\rm K}^2}{K_X}  \right] \,  \frac{a^{-1}}{a_{\rm i}^2} y^2  \delta \, ,
\end{align}
where we used \autoref{eq:nlforcekm} and \autoref{eq:geff_lin_kmou}. $A_{\rm friction}$ is given by \autoref{eq:kmfric}. $w_{\rm eff} = \bar{p}_{\rm eff}/\bar{\rho}_{\rm eff}$ and $\bar{\rho}_{\rm eff}$ are the equation of state and energy density of the effective dark energy fluid component, with $\bar{p}_{\rm eff}$ being the fluid's pressure. These are given in the Einstein frame by \citep{Brax:2014wla,Brax:2014yla}: 
\begin{align}
    \bar{\rho}_{\rm eff} & = - M_{\rm pl}^2 H_0^2 \lambda^2( K - M_{\rm pl}^2 H^2\bar{\phi}'{}^2 K_X) \, ; \\ 
    \bar{p}_{\rm eff} &= M_{\rm pl}^2 H_0^2 \lambda^2 K \, .
\end{align}
We then get 
\begin{equation}
    w_{\rm eff} =  - \frac{K}{K- M_{\rm pl}^2 H^2\bar{\phi}'{}^2 K_X} \, .\label{eq:weffkmouflage}
\end{equation}
We can simplify \autoref{eq:weffvt} further by noting that when adopting the model in \autoref{eq:kmnckinetic} we have
\begin{equation}
    K_X = \frac{1}{H_0^2 \lambda^2 M_{\rm pl}^2 } + K_0 \frac{1}{H_0^{2n} \lambda^{2n} M_{\rm pl}^{2n} } n \, X^{n-1} \, . \label{eq:KXexplicit}
\end{equation}
Substituting this into \autoref{eq:weffkmouflage}, we find the effective dark energy contribution to the potential energy is given by 
\begin{equation}
     \frac{W_{\rm eff}}{E_0} = -\frac{\lambda^2}{3} \left[2 K + \frac{H^2}{H_0^2} \frac{\varphi'{}^2}{\lambda^2} \left(1 + K_0 n X^{n-1} \right)  \right] \frac{a^2}{a_{\rm i}^2} y^2 \, . 
\end{equation}
Finally, we should note that the Newtonian contribution also should have a conformal factor along with $\Omega_{\rm m,0}$
\begin{equation}
\frac{W_{\rm N}}{E_0}  = - A(\varphi) \Omega_{\rm m,0} \frac{a^{-1}}{a_{\rm i}^2} y^2 (1+\delta) \, .  
\end{equation}

\section{Parametrised post-Friedmannian expressions}\label{sec:ppfgeff}

Here we review expressions for the general parametrisation of the effective gravitational constant appearing in the nonlinear Poisson equation as described in Ref.~\cite{Lombriser:2016zfz}. This is based on the parametrised post-Friedmannian framework and is the means of modelling modifications to nonlinear structure formation in the {\tt MG-evolution} code. This parametrisation has also been implemented in the {\tt ReACT} code~\citep{Bose:2022vwi}. 

{\tt MG-evolution} adopts a generalised form of the Vainshtein screening effect given by~\citep{Lombriser:2016zfz}
\be 
\frac{\Delta G_{\text {eff, NL}}}{G} = b\left(\frac{k_*}{k}\right)^{a_f}\left\{\left[1+\left(\frac{k}{k_*}\right)^{a_f}\right]^{1 / b}-1\right\},
\ee 
where NL stands for nonlinear, and $k_*$ and $b$, respectively, characterise the effective screening wavenumber and the interpolation rate between the screened and unscreened regimes. This expression augments the linear theory prediction as given in \autoref{eq:param_mg_evolution} to give the full solution for $G_{\rm eff}$.  We shall briefly provide the particular form of this expression for the three models considered in this work and refer the reader to Ref.~\cite{Lombriser:2016zfz} for more details.

\subsection{nDGP}

To parametrise  nDGP gravity we consider~\citep[see][]{Lombriser:2016zfz}
\be 
\frac{\Delta G_{\rm nDGP, NL}}{G_{\rm N}}=\frac{1}{3 \beta(a)}\left(\frac{k_*}{k}\right)^3\left\{\left[1+\left(\frac{k}{k_*}\right)^3\right]^{\frac{1}{2}}-1\right\},
\ee 
where $k_*$ corresponds approximately to the Vainshtein radius:
\begin{equation}\label{eq:vain_r}
    r_{*} = \frac{16 \pi G_{\rm N} \delta M r_{c}^{2}}{9\beta^{2}},
\end{equation}
where $\delta M$ is the mass enclosed by a spherical region, $r_{c}$ is the crossover scale in nDGP theories, and $\beta(a)$ is given below. The Vainshtein radius effectively defines a region where the fifth force introduced by the scalar field gets shielded. The effective screening wavenumber $k_*$ can in principle be modelled. However, it is treated as a free parameter in  {\tt MG-evolution}. The function $\beta(a)$ reads as
\be
\beta(a) = 1+ 2 H r_c   \Big ( 1+ \frac{\dot{H}}{3 H^2} \Big) \, , 
\ee
and the linear effective gravitational constant in nDGP is  given by
\begin{equation}
    \frac{G_{\rm eff, \ L}}{G_{\rm N}} = 1 + \frac{1}{3\beta(a)}.
\end{equation}
We remind the reader that a cosmological background that matches that of $\Lambda$CDM is assumed.

\subsection{Cubic Galileon} \label{sec:cgextra}

To parametrise the Cubic Galileon we adopt \autoref{eq:param_mg_evolution}, with the nonlinear parametrisation
\be
\frac{\Delta G_{\text {CG, NL}}} {G_{\rm N}} =\left(\frac{k_*}{k}\right)^3\left\{\left[1+\left(\frac{k}{k_*}\right)^3\right]^{\frac{1}{2}}-1\right\} \, .
\ee 
To obtain the linear regime parametrisation we use the effective gravitational potential in Cubic Galileon theory in the linear regime, which reads as \citep{Barreira:2013eea} 
\be
\frac{ \Delta G_{\text {CG, L}}}{G_{\rm N}}=-\frac{2}{3} \frac{c_3 \dot{\phi}^2}{M_{\mathrm{pl}} \mathcal{M}^3 \beta_2} \, ,
\ee
where $\phi$ is the Galileon scalar field. $\beta_2$ and $\mathcal{M}^3$ read,
\begin{align}
\beta_2 &\equiv 2 \frac{\mathcal{M}^3 M_{\mathrm{pl}}}{\dot{\phi}^2} \beta_1, \\
\mathcal{M}^3 & \equiv M_{\mathrm{pl}} H_0^2 \, , 
\end{align}
where
\begin{equation}
    \beta_1 \equiv \frac{1}{6 c_3}\left[-c_2 - \frac{4 c_3}{\mathcal{M}^3}(\ddot{\phi}+2 H \dot{\phi}) + 2 \frac{c_3^2}{M_{\mathrm{pl}}^2 \mathcal{M}^6} \dot{\phi}^4\right] \, . 
\end{equation}
We consider $c_2 = -1$ and we use the tracker solution \citep{Bellini:2017avd},
\be
\xi \equiv \frac{\dot{\phi} H}{M_{\mathrm{pl}} H_0^2}\, , 
\ee
where $\xi$ is a constant and can be written in terms of $c_2, c_3$,
\be
 \xi = -\frac{c_2}{6 c_3 } = \frac{1}{6 c_3 } \, .
\ee 
As a result, we have the following solutions for $\dot \phi$ and $\ddot \phi$
\be
\dot{\phi} =   \frac{\xi M_{\mathrm{pl}} H_0^2}{H}, \; \ddot{\phi} =  - \frac{\xi M_{\mathrm{pl}} H_0^2 \dot{H}}{H^2} \, . 
\ee
Following the discussion presented in Ref.~\cite{Barreira:2013xea} for the background tracker solution, we can derive the Hubble expansion rate as a function of scale factor
\begin{align}
\frac{H^2}{H_0^2} &= \frac{1}{2}\Big[\left(\Omega_{\rm m,0} a^{-3} + \Omega_{\rm r, 0} a^{-4}\right) \\ \nonumber &
+ \sqrt{\left(\Omega_{\rm m,0} a^{-3} + \Omega_{\rm r,0} a^{-4}\right)^2 + 4\left(1 - \Omega_{\rm m,0} - \Omega_{\rm r,0}\right)}\Big] \, , 
\end{align}
where $H_0^2 = \frac{8 \pi G}{3}$ in {\tt MG-evolution} units and $\Omega_{\rm r,0}$ is the radiation energy density fraction today. Computing the cosmic time derivative results in,
\begin{equation}
\begin{aligned}
\frac{\dot{H} +  H^2}{H_0^2} = & -\frac{(a \Omega_{\rm m,0} + 2 \Omega_{\rm r,0})}{4 a^2} \nonumber \\ 
 & - \frac{(a \Omega_{\rm m,0} + \Omega_{\rm r,0}) (3 a \Omega_{\rm m,0} + 4 \Omega_{\rm r,0})}{4 a^6 \sqrt{4 (1 - \Omega_{\rm m,0} - \Omega_{\rm r,0}) + \frac{(a \Omega_{\rm m,0} + \Omega_{\rm r,0})^2}{a^8}}} \\
& + \frac{a^2}{2} \sqrt{4 (1 - \Omega_{\rm m,0} - \Omega_{\rm r,0}) + \frac{(a \Omega_{\rm m,0} + \Omega_{\rm r,0})^2}{a^8}} \, . 
\end{aligned}
\end{equation}

\subsection{K-mouflage}

Here we derive the effective gravitational constant appearing in the nonlinear Poisson equation for the K-mouflage model described in \autoref{sec:kmouflage}. We follow Ref.~\cite{Lombriser:2016zfz}. We also note that a conformal factor, $A(\varphi)$, still needs to be applied to transform the density appearing in the nonlinear Poisson equation which is not included in the $G_{\rm KM,eff}$ expressions below. 

The effective modification assuming a spherically symmetric matter distribution is given as \citep{Winther:2014cia,Brax:2014yla}
\begin{equation}
\frac{\Delta G_{\rm KM, eff}}{G_{\rm N}} = \frac{2 \, \beta_{\rm K}^2}{K_X H_0^2 \lambda^2 M_{\rm pl}^2 } \, . \label{eq:geffkm1}
\end{equation}
Using the Klein-Gordon equation for a spherically symmetric matter distribution we can write $K_X$ as 
\begin{equation}
   K_X^2 X = - \frac{2 \, \beta_{\rm K}^2 }{H_0^4 \lambda^4 M_{\rm pl}^2} F_{\rm N}^2 \, , \label{eq:kmkgeqss}
\end{equation}
where the Newtonian force is just $F_{\rm N} = G_{\rm N} M(<r)/r^2$, $r$ being the physical radial coordinate and $M(<r)$ being the mass enclosed in radius $r$. Substituting for $K_X$ in \autoref{eq:geffkm1} gives 
\begin{equation}
\frac{\Delta G_{\rm KM, eff}}{G_{\rm N}} =\frac{ \beta_{\rm K}  }{M_{\rm pl}}\frac{1}{ F_{\rm N}} \sqrt{-2 X}    \, .  \label{eq:geffkm2}
\end{equation}
Now to solve for $X$ we can adopt the model in \autoref{eq:kmnckinetic}. By using \autoref{eq:KXexplicit} to solve \autoref{eq:kmkgeqss} we get
\begin{equation}
    X = \frac{H_0^2 \lambda^2 M_{\rm pl}^2 }{6 K_0}\frac{\left[ 1-  f(x) \right]^2}{f(x)} \, , \label{eq:solforX}
\end{equation}
where 
\begin{equation}
    f(x) = \left( 1 + x + \sqrt{x (x + 2)} \right)^{\frac{1}{3}} \, , 
\end{equation}
and we have defined $x \equiv  - C_A / r^4 $, $C_A$ being a parameter proportional to $K_0$, defined as 
\begin{equation}
C_A \equiv \frac{54 \beta_{\rm K}^2 G_{\rm N}^2 M^2 }{H_0^2 \lambda^2} K_0 \, . 
\end{equation}
We have written $M= M(<r)$ for compactness. It should be pointed out that $x \in (-2,0)$ yields no solution for $X$ which can be a problem for for very large $r$ and $K_0>0$. This will not generally be an issue as we look for solutions in the nonlinear regime. 

Substituting \autoref{eq:solforX} into \autoref{eq:geffkm2} gives 
\begin{equation}
   \frac{\Delta G_{\rm KM, eff}}{G_{\rm N}} = C_B \frac{1-f(x)}{\sqrt{ x f(x)}} \, ,\label{eq:geffkmfinal}
\end{equation}
where 
\begin{equation}
    C_B \equiv 3 \sqrt{2} \beta_{\rm K}^2 \, . 
\end{equation}
We note that in Ref.~\cite{Lombriser:2016zfz} there seems to be a missing factor of $1/(2 \sqrt{2})$ in Equation.~3.26 in order to have the identification $C_A = C_2^2$. We allow here the case when $x\leq0$ which can occur for $K_0>0$. Further, we note $C_1 = - C_B$, $C_1$ and $C_2$ being the equivalent quantities for $C_A$ and $C_B$ in Ref.~\cite{Lombriser:2016zfz}. We include a \href{https://github.com/nebblu/ACTio-ReACTio/tree/master/notebooks}{\texttt{Mathematica} notebook} with our derivations. 

We now derive the PPF expression from the limits of \autoref{eq:geffkmfinal}:
\begin{align}
\frac{\Delta G_{\rm KM, eff}}{G_{\rm N}}  \rightarrow 2 \beta_{\rm K}^2 \,  \qquad \quad \qquad & {\rm for}  \qquad  |x| \ll 1 \, \,  ({\rm i.e.} \, r^4 \gg |C_A|) \, , \label{eq:gefflimit1} \\ 
\frac{\Delta G_{\rm KM, eff}}{G_{\rm N}}  \rightarrow \frac{C_B r^{4/3}}{(-C_A)^{1/3}} \,  \qquad & {\rm for}  \qquad  |x| \gg 1 \, \,  ({\rm i.e.} \, r^4 \ll |C_A|) \, , \label{eq:gefflimit2}
\end{align}
which are the same limits obtained by Ref.~\cite{Lombriser:2016zfz}.

We now map these onto the (real space) PPF expression, Equation.~5.3 of 
Ref.~\cite{Lombriser:2016zfz} 
\begin{equation}
     \frac{\Delta G_{\rm eff}}{G_{\rm N}} = p_1 p_2 \frac{(1+s^{a_f})^{\frac{1}{p_1}}-1}{s^{a_f}} \, ,  \label{eq:nPPFF0}
\end{equation}
where 
\begin{equation}
    a_f = \frac{p_1}{p_1-1} p_3 \,  , 
\end{equation}
and $s = y_{\rm scr}/y_{\rm h}$. $y$ is the normalised top-hat radius
\begin{equation}
    y \equiv \frac{R_{\rm TH}/a}{R_i/a_i} \, . \label{eq:ydef}
\end{equation}
$R_{\rm TH}$ and $R_i$ are the comoving halo top-hat radius and $a_i$ the initial scale factor. The dimensionless screening scale is given by 
\begin{equation}
 y_{\rm scr} = p_4 a^{p_5} \left(2G_{\rm N}\,H_0 M_{\rm vir}\right)^{p_6} \left(\frac{y_{\rm env}}{y_{\rm h}}\right)^{p_7} \,.
 \label{eq:screeningscale}
\end{equation}
 $y_{\rm env}$ refers to the normalised radius of the environment and $M_{\rm vir}$ is the virial mass of the halo.

 Comparing \autoref{eq:nPPFF0} and \autoref{eq:screeningscale} with \autoref{eq:gefflimit1} and \autoref{eq:gefflimit2} we find, for a choice of $p_1$,
\begin{align}
    p_2 & = \frac{2 \beta_{\rm K}^2}{p_1} ,  \qquad p_3 = \frac{4}{3} \frac{p_1-1}{p_1} , \nonumber \\ 
    p_4 &= \left[\frac{-\sqrt{2} K_0 p_1^3 \beta_{\rm K}^2}{ \lambda^2}\right]^{\frac{1}{4}} \Omega_{\rm m,0}^{\frac{1}{3}}  , \nonumber \\  p_5 &  = -1 , \qquad  p_6 = 1/6, 
   \qquad p_7  = 0 \, . \label{eq:dgppi}
\end{align}
We note that whether there's a $p_1$ in $p_3$ depends on whether $p_1$ is positive or negative (see Equation.~2.14 of Ref.~\cite{Lombriser:2016zfz}). 
We have also used $M_{\rm vir} \approx 4 \pi \Omega_{\rm m,0} \rho_{\rm crit} R_{\rm th}^3 /3 $. 

\bsp	
\label{lastpage}
\end{document}